\documentclass[aps,pre,twocolumn,showpacs]{revtex4}
\usepackage{graphicx}
\usepackage{amsmath,amscd,amssymb}
\usepackage{color}

\newcommand{\be}{\begin{equation}}
\newcommand{\ee}{\end{equation}}
\newcommand{\bea}{\begin{eqnarray}}
\newcommand{\eea}{\end{eqnarray}}
\newcommand{\lan}{\left\langle}
\newcommand{\ran}{\right\rangle}
\newcommand{\br}{\mathbf{r}}

\newcommand{\bx}{\mathbf{x}}

\newcommand{\e}{\varepsilon}

\begin{document}

\title{Excluded volume effects in macromolecular forces and ion-interface interactions}

\author{Sahin Buyukdagli$^{1}$\footnote{email:~\texttt{sahin\_buyukdagli@yahoo.fr}} and T. Ala-Nissila$^{1,2}$\footnote{email:~\texttt{Tapio.Ala-Nissila@aalto.fi}}}
\affiliation{$^{1}$Department of Applied Physics and COMP center of Excellence, Aalto University School of Science, P.O. Box 11000, FI-00076 Aalto, Espoo, Finland\\
$^{2}$Department of Physics, Brown University, Providence, Box 1843, RI 02912-1843, U.S.A.}
\date{\today}

\begin{abstract}
A charged Yukawa liquid confined in a slit nanopore is studied in order to understand excluded volume effects in the interaction force between the pore walls. A previously developed self-consistent scheme (S. Buyukdagli el al., J. Stat. Mech. P05033 (2011)) and a new simpler variational procedure that self-consistently couple image forces, surface charge induced electric field and pore modified core interactions are used to this aim. For neutral pores, it is shown that with increasing pore size, the theory predicts a transition of the interplate pressure from an attractive to a strongly repulsive regime associated with an ionic packing state, an effect observed in previous Monte Carlo simulations for hard core charges. We also establish the mean-field theory of the model and show that for dielectrically homogeneous pores, the mean-field regime of the interaction between the walls corresponds to large pores of size $d>4$ {\AA}. The role of the range of core interactions in the ionic rejection and interplate pressure is thoroughly analyzed. We show that the physics of the system can be split into two screening regimes. The ionic packing effect takes place in the regime of moderately screened core interactions characterized with the bare screening parameter of the Yukawa potential $b\lesssim 3/\ell_B$, where $\ell_B$ is the Bjerrum length. In the second regime of strongly screened core interactions $b\gtrsim 3/\ell_B$, solvation forces associated with these interactions positively contribute to the ionic rejection driven by electrostatic forces and enhance the magnitude of the attractive pressure. For weakly charged pores without a dielectric discontinuity, core interactions make a net repulsive contribution to the interplate force and also result in oscillatory pressure curves, whereas for intermediate surface charges, these interactions exclusively strengthen the external pressure, thereby reducing the magnitude of the net repulsive interplate force. The pronounced dependence of the interplate pressure and ionic partition coefficients on the magnitude and the range of core interactions indicates excluded volume effects as an important ion specificity and a non-negligible ingredient for the stability of macromolecules in electrolyte solutions.

\end{abstract}

\maketitle

\section{Introduction}

The first experimental test of the Lifshitz theory of Van der Waals (vdW) forces between surfaces was performed for surface separations between 100 to 1000 nm by Derjaguin and collaborators in 1954~\cite{Derja}. The improvement of experimental technics allowed in 1970's the confirmation of the theory for interplate separations of a few nanometers~\cite{Tabor1,Tabor2}. These experimental breakthroughs were followed by a simpler reformulation of the Lifshitz theory~\cite{Ninh,Hough}. It is known that the stability of various biological and chemical systems such as membrane assemblies~\cite{Dub}, colloidal suspensions~\cite{Holm}  or cement paste~\cite{Jonsson} result mainly from the competition between the attractive vdW forces and repulsive double layer interactions induced by the charge groups at the surface of the molecules. This competition is the basis of the Derjaguin-Landau-Verwey-Overbeek (DLVO) theory~\cite{DLVO,Isr}. Although the DLVO theory has been successful to explain several phenomena in colloidal science, it treats these opposing forces in an additive way. The additivity assumption is, however, an uncontrolled approximation that is expected to break down at ionic concentrations where many-body effects become significant.

A gaussian field theory of heterogeneous ionic solutions that couples both effects in a systematic way was proposed in Ref.~\cite{PodWKB}. By comparison with MC simulations, it was shown that the theory reproduces the correct trend for the deviation of the exact interplate pressure from the mean-field (MF) prediction. During the last decade, the field theoretical formulation of heterogeneous ionic liquids has been applied at the gaussian level to more complicated systems in order to understand the impact of various effects on macromolecular forces, such as surface charging~\cite{David4}, charge disorder~\cite{PodAli}, surface polarity~\cite{David2}, and dielectric disorder~\cite{David1}. An analytical theory for the non-equilibrium behaviour of Casimir forces has been also proposed in Ref.~\cite{David5}. By construction, the gaussian theories are known to be valid for dilute ionic concentrations. A first order cumulant expansion that goes beyond the gaussian theory and allows to consider higher concentration regimes was introduced in Ref.~\cite{David3}. In addition to the screened vdW forces, this calculation was shown to give rise to a new attractive force, namely a depletion force that originates from the density difference between the bulk electrolyte and the interior of the pore.

Close to molar concentrations where various electrostatic effects act in a self-consistent way, one can exclusively rely on non-perturbative methods. Electrostatic self-consistent equations for the two point correlation function and the surface charge induced electrostatic potential were first derived within a variational procedure in Ref.~\cite{netzvar}. These equations were recently shown to be equivalent to Hartree equations that can also be obtained from a summation over a particular class of perturbative diagrams~\cite{Lee}. Within a Wentzel-Kramers-Brillouin (WKB) approximation, they were solved for cylindrical systems with a dielectric discontinuity in order to understand ionic correlation effects. A variational calculation based on a more restricted variational kernel that can self-consistently take into account the depletion forces discussed above was also proposed for neutral slit pores in Ref.~\cite{hatlo}. A different variational approach based on a separation of the Coulomb potential into a short and a long wavelength component was presented for charged slit pores without a dielectric discontinuity in Ref.~\cite{hatloSM} and in the presence of a dielectric discontinuity but without salt in Ref.~\cite{hatloEPL}. By comparison with MC simulations, it was shown that the predictions of the theory for ionic densities and interplate pressure were very accurate from weak to strong coupling limit. In order to be able to consider on the same footing the salt, the dielectric discontinuity and the surface charge effects, we proposed in Ref.~\cite{PRE} a simpler variational approach for electrolytes confined in slit pores. The approach in question is inspired by a modified Onsager-Samaras (OS) approach frequently used in nanofiltration studies~\cite{yarosch,YarII,Szymczyk} and it considers uniform trial screening parameters whose value can differ from the bulk one due to confinement effects. The variational scheme was shown to agree well with MC simulation results beyond the MF regime. It was also shown that the approach applied to cylindrical ion channels yields a new type of liquid-vapor phase transition that we proposed as the underlying mechanism behind the ionic current fluctuations observed in experiments~\cite{PRL,JCP}.

The field theoretic approaches discussed above do not consider excluded volume effects associated with core-core collisions between the charges. These effects included in numerical simulations of electrolyte solutions are known to be non-negligible if the packing fraction of the electrolyte becomes important. More precisely, Monte Carlo (MC) simulation results of ions with hard core (HC) interactions show that for an electrolyte with a bulk density about 1 M, ions close to the solid interface feel an attraction towards the wall~\cite{MCwet1,MCwet2}. Previous integral theories of HC charges at planar surfaces showed that this feature is a wetting effect caused by the particle collisions within the bulk that push ions towards the interfaces~\cite{kjal1,kjal2,kjal3,kjal4}. We have recently developed a self-consistent calculation scheme~\cite{jstat} in order to study a field theoretic model of charged liquids with repulsive Yukawa interactions~\cite{DunYuk}. It was shown that the theory is able to reproduce the ionic wetting effect in question and exhibits a good agreement with MC simulation results for the density profile of neutral Yukawa particles at simple interfaces. We also investigated ion size effects on the adsorption of ions onto dielectric interfaces as well as on the dielectric exclusion mechanism  from slit nanopores. It is important to emphasize at this stage that although different potentials could be used in order to model core-core interactions, our choice of a Yukawa potential is motivated by the fact that its inverse is well defined. We note that similar charged liquids with repulsive Yukawa interactions has been studied in bulk systems with MC simulations in order to investigate vapor-liquid equilibrium~\cite{Kris}. Furthermore, the most important benefits of the present theory over MC simulations are the transparency of the closure equations that allows an easy interpretation of the underlying physics, and a considerable reduction of the computation time. Indeed, in the most complicated case of the charged Yukawa liquid confined in a dielectrically heterogeneous pore with charged walls, the numerical solution of the general self consistent equations (GVS-see below) does not exceed 10 minutes for pores of total thickness $d<3$ nm.

A different effect related to this wetting phenomenon was observed in MC simulations of HC charges confined in slit nanopores without dielectric discontinuity~\cite{Bratko1,Bratko2}. For large bulk concentrations, the interplate pressure was shown to interpolate between an attractive regime of ionic depletion at small pore sizes and a repulsive regime characterized by an ionic packing state at large interplate separations. Although the first regime associated with confinement effects is included in field theoretic models of vdW interactions~\cite{netzvdw,hatlo}, the second regime of ionic packing driven by core-core collisions is absent in these theories. In this article, we revisit the theory developed in Ref.~\cite{jstat} in order to show that the model includes both regimes and also thoroughly analyze the underlying mechanism responsible for the interpolation between them.

\begin{figure}
\includegraphics[width=1.3\linewidth]{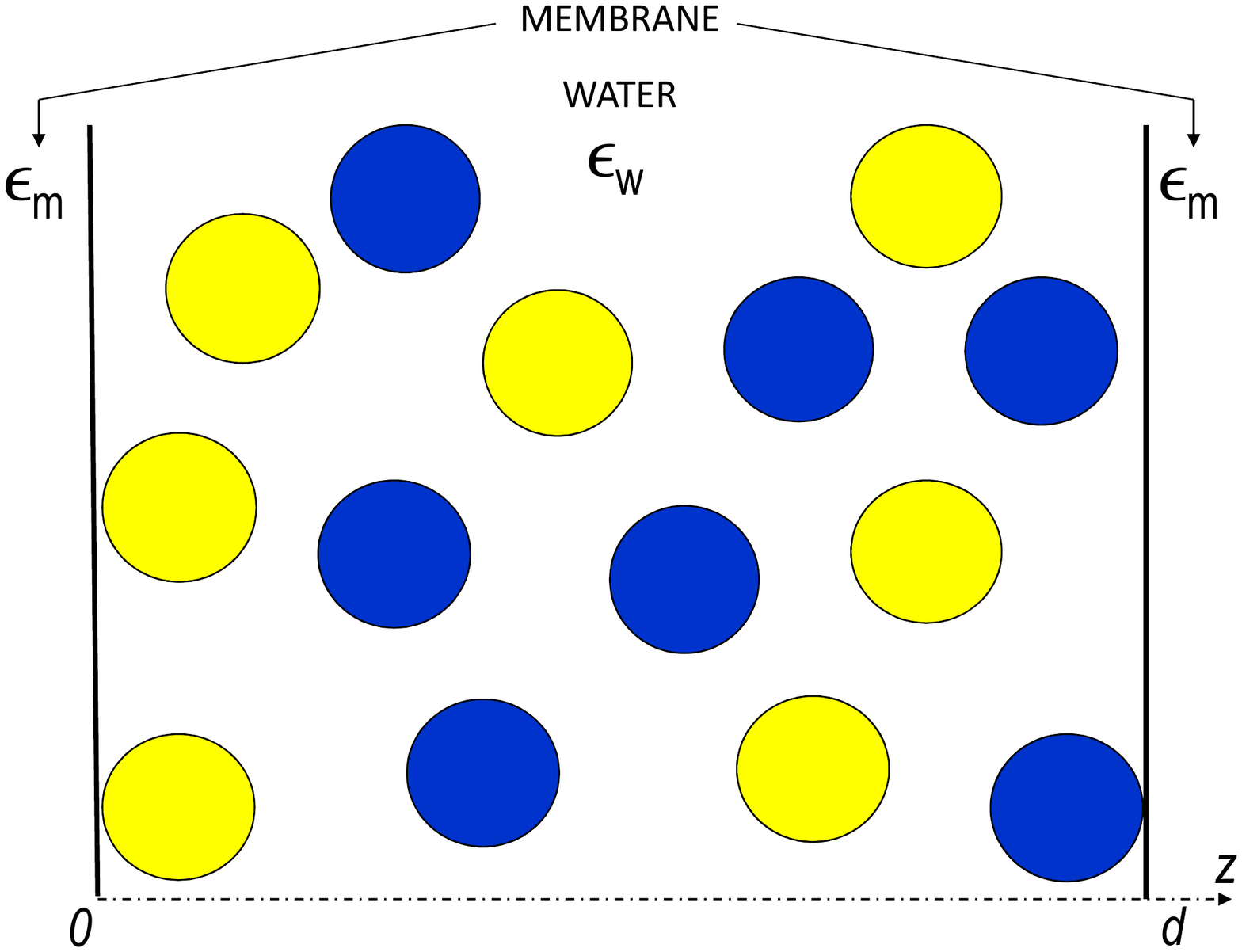}
\caption{(Color online) Geometry for a slit-like pore of thickness $d$. The dielectric permittivities of the pore and the membrane are respectively $\e_w$ and $\e_m$.}
\label{sketch}
\end{figure}

The article is organized as follows. We explain in section 2 the derivation of the field theoretic model for the charged Yukawa system confined between two membrane walls containing fixed surface charges with a uniform amplitude $\sigma_s$ (see Fig.~\ref{sketch}). Each wall separates two dielectric media, namely the membrane matrix composed of biological or synthetic substance associated with a low dielectric permittivity ($\e_m=2$), and the pore medium that contains the solvent molecules (i.e. water) and solvated ions. \textcolor{black}{The pore is in contact with an external particle reservoir at the extremities}, and the electrostatic interactions between the ions \textcolor{black} {in the bulk reservoir} are modeled with a Coulomb potential $v_c=\ell_B/r$, where $\ell_B$ is the Bjerrum length (defined in section 2). The repulsive core-core interactions resulting from the excluded volume associated with the size of hydrated ions is taken into account with a Yukawa potential of the form $w(r)=\ell_ye^{-br}/r$, where the model parameters $\ell_y$ and $b$ respectively fixe the amplitude and the range of these interactions. \textcolor{black}{We note that a mapping between these parameters and the effective ion radius has been presented in Ref.~\cite{jstat}. Since we aim in the present work at understanding the importance of core-core interactions with respect to the magnitude of electrostatic interactions, the parameters $\ell_y$ and $b$ will be varied in terms of the Bjerrum length. Furthermore,} the solvent molecules renormalize the dielectric permittivity of the air to a high value ($\e_w=78$) and in the presence of ions, the resulting dielectric discontinuity between the membrane and the pore media gives rise to induced polarization charges that are called \textit{image charges}. \textcolor{black}{Although an ion located close to a single dielectric interface has a single image charge, the confinement of the ion between two interfaces gives rise to an infinite number of images.} The present theory can fully take into account the interaction between an ion in the pore and its multiple images, as well as the screening of \textcolor{black} {these} image interactions by the surrounding ions in a consistent way.  Section 3 is devoted to the derivation of the computational schemes. We first develop the MF theory of the model and calculate the MF pressure for the slit system. Then, we introduce two self-consistent calculation schemes that account for the correlation effects neglected at the MF level. Namely, we revisit the derivation of the variational equations introduced in Ref.~\cite{jstat} and introduce as well a simpler variational approach. We also derive the interaction force between the pore walls within these self-consistent methods. The numerical results are discussed in Section 3. In the first part, we apply the theory to neutral Yukawa particles confined in the slit pore in order to evaluate the net contribution from excluded volume effects to the interplate pressure. The second part considers the interplay between electrostatic and core interactions for neutral pores with various matrix permittivities. In the third part, we thoroughly analyze the impact of the range of core interactions on ionic rejection rates and the interaction force between the plates. Finally, we discuss in the fourth part the effect of core interactions on the interplate pressure in the presence of a fixed surface charge. In all cases considered above, we compare the restricted variational scheme with the general one and also illustrate the MF predictions in order to identify the MF regime of the theory. The limitations of the theory, potential generalizations and applications are discussed in the conclusion.

\section{Field theoretic model for the charged Yukawa fluid}

We review in this section the derivation of the grand canonical partition function of the charged Yukawa model introduced in Ref.~\cite{DunYuk} that will be the starting point for the following sections. The canonical partition function of interacting charged Yukawa particles reads
\be\label{PartCan}
Z_c=\prod_{i=1}^p\frac{e^{N_iE_s}}{N_i!\lambda_T^{3N_i}}\int\prod_{j=1}^{N_i}\mathrm{d}\bx_{ij}e^{-H_c\left(\{\bx_{ij}\}\right)
-H_y\left(\{\bx_{ij}\}\right)}
\ee
where $p$ is the number of particle species, $N_i$ is the number of particles for each species, and $\lambda_T$ is the thermal wavelength of each particle. The electrostatic and repulsive core interactions are given by
\bea
\label{Hc}
H_{c}\left(\{\bx_{ij}\}\right)&=&\frac{1}{2}\int\mathrm{d}\br\mathrm{d}\br'\rho_c(\br)v_c(\br,\br')\rho_c(\br')\\
\label{Hp}
H_{y}\left(\{\bx_{ij}\}\right)&=&\frac{1}{2}\int\mathrm{d}\br\mathrm{d}\br'\rho_p(\br)w(\br,\br')\rho_p(\br')\\
&&+\int\mathrm{d}\br V_w(\br)\rho_p(\br),\nonumber
\eea
where $\rho_p(\br)=\sum_{i=1}^p\sum_{j=1}^{N_i}\delta(\br-\bx_{ij})$ is the particle density, \textcolor{black}{$x_{ij}$ is the coordinate of the particle $j$ of species $i$}, $\rho_c(\br)=\sum_{i=1}^p\sum_{j=1}^{N_i}q_i\delta(\br-\bx_{ij})+\sigma(\br)$ the total charge density, $q_i$ the valency of mobile ions and $\sigma(\br)$ stands for a uniform fixed surface charge density (expressed in units of the elementary charge $e$). We note that in this article, we will exclusively consider the case of negatively charged membrane pores, i.e. $\sigma(\br)<0$. \textcolor{black}{Furthermore, the wall potential $V_w(\br)$ takes into account the fact that the particles cannot penetrate into the membrane} by restricting the phase space accessible to the ions according to
\bea\label{StPot}
&&V_w(z)=0\hspace{0.5mm}\mbox{,}\hspace{4mm}0\leq z\leq d\nonumber\\
&&V_w(z)=\infty\hspace{0.5mm}\mbox{,}\hspace{4mm}z<0\hspace{1mm}\mbox{and}\hspace{2mm}z>d.
\eea
Moreover, the electrostatic and core potentials are respectively defined as the inverse of the following operators
\bea\label{coulomb}
v_c^{-1}(\br,\br')&=&-\frac{k_BT}{e^2}\nabla\left[\e(\br)\nabla\delta(\br-\br')\right]\\
\label{Yukawa}
w^{-1}(\br,\br')&=&\frac{b^2-\Delta}{4\pi\ell_y}\delta(\br-\br'),
\eea
where we introduced the spatially varying dielectric permittivity $\e(\br)$. In the case of a slit pore that confines a solvent composed of water molecules, the permittivity is given by $\e(z)=\e_w\theta(z)\theta(d-z)+\e_m[\theta(-z)+\theta(z-d)]$, where $\e_m$ and $\e_w$ are respectively the dielectric permittivity of the membrane and the water medium. We first note that the Yukawa operator Eq.~(\ref{Yukawa}) can be easily inverted in Fourier space, which yields the short range core interaction potential in the form $w(\br)=\ell_ye^{-b|\br|}/|\br|$. Furthermore, the self energy of ions that should be subtracted from the total Hamiltonian is given by $E_s=\frac{q_i^2}{2}v_c^b(\br-\br')|_{\br=\br'}+\frac{1}{2}w(\br-\br')|_{\br=\br'}$, with the Coulomb operator in a bulk medium defined as ${v^b_c}^{-1}(\br,\br')=-\frac{k_BT\e_w}{e^2}\Delta\delta(\br-\br')$, where $\ell_B=e^2/(4\pi\e_wk_BT)\simeq7$ {\AA} is the Bjerrum length at ambient temperature $T=300$ K. The inverse of the Coulomb kernel is the Coulomb potential $v^b_c(\br)=\ell_B/|\br|$.

Performing two Hubbard-Stratanovitch transformations in order to pass from the density to the field representation and using the relation $Z_G=\prod_{i=1}^p\sum_{N_i\geq0}e^{\mu_i N_i}Z_c$, one can obtain the Grand canonical partition function from the canonical one in the form
\be\label{Zg}
Z_G=\int \mathcal{D}\phi\mathcal{D}\psi\;e^{-H[\phi,\psi]}
\ee
where the functional Hamiltonian reads
\bea\label{HamFunc}
H[\phi,\psi]&=&\int
\mathrm{d}\br\left[\frac{\left[\nabla\phi(\br)\right]^2}{8\pi\ell_B(\br)}-i\sigma(\br)\phi(\br)\right]\\
&&+\int\frac{\mathrm{d}\br}{8\pi\ell_y}\left[\left[\nabla\psi(\br)\right]^2+b^2\psi^2(\br)\right]\nonumber\\
&&-\sum_i\lambda_i \int\mathrm{d}\br e^{E_s-V_w(\br)+i \left[q_i\phi(\br)+\psi(\br)\right]}\nonumber.
\eea
In Eq.~(\ref{HamFunc}), $\phi(\br)$ is the fluctuating electrostatic potential and $\psi(\br)$ stands for the Yukawa potential associated with core interactions between the particles. We also introduced above the spatially varying Bjerrum length $\ell_B(\br)=e^2/\left[4\pi\e(\br)k_BT\right]$ and the rescaled particle fugacity $\lambda_i=e^{\mu_i}/\lambda_T^3$. We finally note that in this article, we will consider exclusively the case of symmetric electrolytes with bulk concentration $\rho_b^+=\rho_b^-=\rho_b$ and valency $q_+=-q_-=q$, where $+$ is for cations and $-$ for anions.

\section{Calculation schemes}
\subsection{MF theory}
\subsubsection{MF equations}

The MF limit of the partition function Eq.~(\ref{Zg}) corresponds to the parameter regime associated with a low surface charge and a dilute electrolyte. The physics of the liquid is described in this regime by the MF equations that follow from the saddle-point evaluation of the functional integral Eq.~(\ref{Zg}), that is $\delta H/\delta\phi(\br)=0$ and $\delta H/\delta\psi(\br)=0$~\cite{DunYuk}, which yields
\bea
\label{MFphi}
&&\Delta\phi(\br)-\kappa_{DH}^2e^{-V_w(\br)-\psi(\br)+\psi_b}\sinh\phi(\br)=-4\pi q\ell_B\sigma(\br)\nonumber\\
&&\\
\label{MFpsi}
&&\Delta\psi(\br)-b^2\psi(\br)+8\pi\ell_y\rho_be^{-V_w(\br)-\psi(\br)+\psi_b}\cosh\phi(\br)=0,\nonumber\\
&&
\eea
where we rescaled the electrostatic and Yukawa fields according to $\bar\psi(\br)=-i\psi(\br)$ and $\bar\phi(\br)=-iq\phi(\br)$, and dropped the bar over the potentials for the sake of simplicity. In the equations above, we introduced the Debye-H\"{u}ckel screening parameter $\kappa_{DH}^2=8\pi\ell_Bq^2\rho_b$. Furthermore, the particle fugacity within the pore was determined from the chemical equilibrium condition between the pore and the bulk reservoir, i.e. $\lambda_i=\lambda_{b,i}=\rho_be^{\psi_b}$, where
\be
\psi_b=\frac{8\pi\ell_y\rho_b}{b^2}
\ee
stands for the bulk limit of the Yukawa potential~\cite{jstat}.

The boundary conditions associated with Eqs.~(\ref{MFphi}) and~(\ref{MFpsi}) that was derived in Ref.~\cite{jstat} for a charge distribution of the form $\sigma(z)=\sigma\left[\delta(z)+\delta(z-d)\right]$ read
\bea\label{boundphi2}
&&\left.\frac{d\phi}{dz}\right|_{z=0^+}=-4\pi q\ell_B\sigma\\
\label{boundpsi2}
&&\left.\frac{d\psi}{dz}\right|_{z=0^+}=b\psi\left(z=0^+\right)\\
\label{boundphi3}
&&\left.\frac{d\phi}{dz}\right|_{z=d^-}=4\pi q\ell_B\sigma\\
\label{boundpsi3}
&&\left.\frac{d\psi}{dz}\right|_{z=d^-}=-b\psi\left(z=d^-\right).
\eea
In this article, we will consider the case of negatively charged pores, i.e. $\sigma=-\sigma_s$, with $\sigma_s\geq0$.

We note that the MF equations~(\ref{MFphi}) and~(\ref{MFpsi}) do not contain an analytical solution. However, they can be linearized in the regime $\psi_b<1$ and $\kappa_{DH}\mu>1$, where $\mu=1/(2\pi q\ell_B\sigma_s)$ stands for the Gouy-Chapman length. The linearization gives
\bea
&&\frac{d^2\phi}{dz^2}-\kappa_{DH}^2\phi=-4\pi q\ell_B\sigma(z)\\
&&\frac{d^2\psi}{dz^2}-\kappa_{yb}^2\psi=-\kappa_{yb}^2\psi_b,
\eea
where $\kappa_{yb}^2=b^2+8\pi\ell_y\rho_b$. These simple differential equations can be solved with the boundary conditions~(\ref{boundphi2})-(\ref{boundpsi3}). One obtains
\be\label{solMF2in}
\psi(z)=\psi_b-\frac{b\psi_b\cosh\left[\kappa_{yb}\left(d/2-z\right)\right]}
{\kappa_{yb}\sinh\left[\kappa_{yb}d/2\right]+b\cosh\left[\kappa_{yb}d/2\right]}.
\ee
for the Yukawa field and
\be\label{solMF2in}
\phi(z)=-\frac{2}{\kappa_{DH}\mu}\frac{\cosh\left[\kappa_{DH}\left(d/2-z\right)\right]}{\sinh\left[\kappa_{DH}d/2\right]}
\ee
for the electrostatic field.

\subsubsection{MF pressure}

The equations derived so far were already introduced in Ref.~\cite{jstat}. In this part, we will use these results in order to compute the MF level interplate pressure. This can be obtained either from the derivative of the MF free energy with respect to $d$, or by finding the constant of integration associated with the equations~(\ref{MFphi}) and~(\ref{MFpsi}). By multiplying Eqs.~(\ref{MFphi}) and~(\ref{MFpsi}) respectively with $\phi'(z)$ and $\psi'(z)$, and integrating once with respect to $z$, one obtains after an integration by part
\bea
&&\frac{\phi'^2}{2}-\kappa_{DH}^2e^{\psi_b-\psi}\cosh\phi-\kappa_{DH}^2\int\mathrm{d}ze^{\psi_b-\psi}\psi'\cosh\phi=c_1\nonumber\\
&&\frac{\psi'^2}{2}-\frac{b^2}{2}\psi^2+8\pi\ell_y\rho_b\int\mathrm{d}ze^{\psi_b-\psi}\psi'\cosh\phi=c_2.
\eea
By combining these two relations in order to cancel the integral terms on the lhs, one obtains the first integral for the system of Eqs.~(\ref{MFphi}) and~(\ref{MFpsi}), which can be in turn related to the interplate pressure. The latter reads
\bea\label{MFPr1}
\beta\Pi=&-&\frac{1}{8\pi\ell_y}\left[\psi'^2(z)-b^2\left(\psi^2(z)-\psi_b^2\right)\right]-\frac{\phi'^2(z)}{8\pi\ell_Bq^2}\nonumber\\
&+&\rho_+(z)+\rho_-(z)-2\rho_b,
\eea
where we subtracted the bulk osmotic pressure $\Pi_b=2\rho_b+b^2\psi_b^2/(8\pi\ell_y)$ and introduced the MF level local densities in the form
\be
\rho_{\pm}(z)=\rho_be^{\psi_b-\psi(z)\mp\phi(z)}.
\ee
As implied by the mechanical equilibrium, $\Pi$ does not depend on the coordinate $z$. The evaluation of the rhs of Eq.~(\ref{MFPr1}) at $z=d/2$ yields
\be\label{MFPr3}
\beta\Pi=\frac{b^2}{8\pi\ell_y}\left[\psi^2(d/2)-\psi_b^2\right]
+\rho_+(d/2)+\rho_-(d/2)-2\rho_b.
\ee
This relation indicates that at the MF level, the net pressure is the osmotic pressure difference between the mid-pore and the bulk reservoir. Moreover, by setting on the rhs of Eq.~(\ref{MFPr1}) $z=d$, one can relate the interplate pressure to the contact ion density as
\be
\label{MFPr2}
\beta\Pi=-\psi_b\rho_b-2\pi\ell_B\sigma_s^2+\rho_+(d)+\rho_-(d)-2\rho_b.
\ee
Eq.~(\ref{MFPr2}) is a contact value relation modified by core interactions. By taking the limit $d\to\infty$ where the net pressure vanishes, one obtains for the single interface system a modified \textit{Grahame equation} that relates the total contact density of ions to the physical parameters of the system,
\be\label{Grahame}
\rho_s^++\rho_s^--2\rho_b=\psi_b\rho_b+2\pi\ell_B\sigma_s^2.
\ee
It is seen in this relation that core interactions increase the total particle density on the wall. We note that the same effect is responsible for the adsorption of HC charges onto neutral interfaces in MC simulations~\cite{MCwet1,MCwet2}.  The corresponding wetting mechanism driven by core collisions was investigated in detail in Ref.~\cite{jstat}. Furthermore, by substituting Eq.~(\ref{Grahame}) into Eq.~(\ref{MFPr2}), one gets
\be\label{MFPr3}
\beta\Pi=\rho_+(d)+\rho_-(d)-\rho_s^+-\rho_s^-.
\ee
Eq.~(\ref{MFPr3}) shows that as in the case of the primitive ion model, the net pressure is also equal to the variation in the total contact density as one approaches the plates from an infinite to a finite separation $d$.

In the linear limit, the electrostatic and core contributions decouple, and one gets from Eq.~(\ref{MFPr1}) $\Pi\simeq\Pi_{core}+\Pi_{el}$, where the part associated with excluded volume effects reads
\be\label{lincore}
\beta\Pi_{core}=\frac{\psi_b\rho_b\kappa_{yb}^2}{\left[\kappa_{yb}\sinh\left(\kappa_{yb}d/2\right)+b\cosh\left(\kappa_{yb}d/2\right)\right]^2},
\ee
and the electrostatic pressure is given by~\cite{DH}
\be
\beta\Pi_{el}=\frac{4\rho_b}{\mu^2\kappa_{DH}^2\sinh^2\left(\kappa_{DH}d/2\right)}.
\ee
For large interplate separations, Eq.~(\ref{lincore}) reduces to
\be
\beta\Pi_{core}\simeq\frac{4\psi_b\rho_b\kappa_{yb}^2}{(\kappa_{yb}+b)^2}e^{-\kappa_{yb}d}.
\ee
Hence, in this limit, the MF level interplate force induced by core collisions exhibits an exponential decay characterized by the length scale $\kappa_{yb}$. The MF results of Eqs.~(\ref{MFPr1}) and~(\ref{lincore}) will be compared in section IV with the self consistent results that will be derived below.

\subsection{Self-consistent approaches}
\subsubsection{General variational scheme}

A self-consistent calculation scheme that allows to partially capture correlation effects neglected at the MF level was introduced in Ref.~\cite{jstat}. The variational reference Hamiltonian of this self-consistent approach is a Gaussian functional in Yukawa and electrostatic potentials, and it is of the form $H_0=H_{0\phi}+H_{0\psi}$, with the Coulombic and Yukawa parts respectively given by
\bea
&&H_{0\phi}=\frac{1}{2}\int_{\br,\br'}\left[\phi(\br)-i\phi_0(\br)\right]
v^{-1}_0(\br,\br')\left[\phi(\br')-i\phi_0(\br')\right]\nonumber\\
&&H_{0\psi}=\frac{1}{2}\int_{\br,\br'}\left[\psi(\br)-i\psi_0(\br)\right]
w^{-1}_0(\br,\br')\left[\psi(\br')-i\psi_0(\br')\right],\nonumber\\
\label{H0phi}
\eea
where the variational electrostatic and Yukawa kernels are given by
\bea\label{DH1}
&&v_0^{-1}(\br,\br')=\frac{k_BT}{e^2}\left[-\nabla(\e(\br)\nabla)+\e(\br)\kappa_c^2(\br)\right]\delta(\br-\br')\nonumber\\
&&\\
\label{DH2}
&&w_0^{-1}(\br,\br')=\frac{-\Delta+\kappa_y^2(\br)}{4\pi\ell_y}\delta(\br-\br').
\eea
The piecewise trial screening parameters introduced in Eqs.~(\ref{DH1}) and~(\ref{DH2}) read $\kappa_c(z)=\kappa_c\theta(z)\theta(d-z)$ and $\kappa_y(z)=b[\theta(-z)+\theta(z-d)]+\kappa_y\theta(z)\theta(d-z)$. The variational Grand potential to be optimized with respect to the trial functions $\kappa_y$, $\kappa_c$, $\psi_0(z)$, and $\phi_0(z)$, is defined as $\Omega_v=\Omega_0+\lan H-H_0\ran_0$, where $\Omega_0=-\ln Z_0=-\ln\int \mathcal{D}\phi\mathcal{D}\psi\;e^{-H_0[\phi,\psi]}=\Omega_{0\phi}+\Omega_{0\psi}$ is the gaussian part that in the absence of core interactions yields the vdW interaction energy~\cite{netzvdw}. Evaluating the functional integrals in $\Omega_v$, the variational Grand potential takes the form
\bea\label{grandPot}
\Omega_v&=&\Omega_{0\phi}+\Omega_{0\psi}
+S\int\mathrm{d}z\left(-\frac{\left[\nabla\phi_0(z)\right]^2}{8\pi\ell_B(z)}+\sigma(z)\phi_0(z)\right)\nonumber\\
&&-S\int\frac{\mathrm{d}z}{8\pi\ell_y}\left(\left[\nabla\psi_0(z)\right]^2+b^2\psi_0^2(z)\right)\nonumber\\
&&-S\int_0^d\mathrm{d}z\left[\frac{\kappa_c^2}{8\pi\ell_B}v_0(\br,\br)
+\frac{\kappa_y^2-b^2}{8\pi\ell_y}w_0(\br,\br)\right]\nonumber\\
&&-S\sum_i\int_0^d\mathrm{d}z\rho_i(z).
\eea
In Eq.~(\ref{grandPot}), the one loop contributions read
\bea\label{GausCoPhi}
&&\Omega_{0\phi}=-\ln\int \mathcal{D}\phi\;e^{-H_{0\phi}[\phi]}\\
\label{GausCoPsi} &&\Omega_{0\psi}=-\ln\int
\mathcal{D}\psi\;e^{-H_{0\psi}[\psi]} \eea
and the local density is given by
\bea\label{LocRef1}
\rho_i(\br)=\lambda_ie^{-V_w(\br)-\frac{q_i^2}{2}\left[v_0(\br,\br)-v_c^b(0)\right]-\frac{1}{2}\left[w_0(\br,\br)-w(0)\right]}\nonumber\\
\times e^{-\psi_0(\br)-q_i\phi_0(\br)}.
\eea
We now introduce the external potentials
\bea\label{PotEl}
&&V_c(z)=\frac{q^2}{2}\left[\ell_B(\kappa_{DH}-\kappa_c)+\delta v_0(z)\right]\\
\label{PotYu}
&&V_y(z)=\frac{1}{2}\left[\ell_y(\kappa_{yb}-\kappa_y)+\delta
w_0(z)\right].
\eea
The potentials $V_c(z)$ and $V_y(z)$ were computed in Ref.\cite{jstat} by inverting the operators~(\ref{DH1}) and~(\ref{DH2}). The derivation is briefly explained in Appendix~\ref{appendixCoef}. The result reads
\bea\label{PMFslit1}
V_c(z)&=&\frac{q^2\ell_B}{2}(\kappa_{DH}-\kappa_c)\\
&+&\frac{q^2\ell_B}{2}\int_0^\infty\frac{\mathrm{d}kk\Delta_c}{\rho_c}\frac{e^{-2\rho_cz}
+e^{-2\rho_c(d-z)}+2\Delta_c e^{-2\rho_cd}}{1-\Delta_c^2e^{-2\rho_cd}}\nonumber\\
\label{PMFslit2}
V_y(z)&=&\frac{\ell_y}{2}(\kappa_{yb}-\kappa_y)\\
&+&\frac{\ell_y}{2}\int_0^\infty\frac{\mathrm{d}kk\Delta_y}{\rho_y}\frac{e^{-2\rho_yz}
+e^{-2\rho_y(d-z)}+2\Delta_y e^{-2\rho_yd}}{1-\Delta_y^2e^{-2\rho_yd}}.\nonumber
\eea
The potentials $V_c(z)$ and $V_y(z)$ generate purely repulsive forces that exclude ions from simple interfaces and pores~\cite{jstat}. $V_c(z)$ contains image charge forces associated with the dielectric discontinuity between the pore and the membrane as well as electrostatic solvation forces that originates from the distortion of the ionic cloud around a central charge by the pore walls. $V_y(z)$ contains solvation forces associated with the modification of the screening of core interactions by the interfaces.

By rescaling the electrostatic potential according to $\bar\phi_0(z)=q\phi_0(z)$, the density function expressed in terms of the potentials $V_c(z)$ and $V_y(z)$ takes the form
\be\label{locden}
\rho_\pm(z)=\rho_be^{-V_w(z)-V_c(z)-V_y(z)+\psi_b-\psi_0(z)\mp\bar\phi_0(z)}.
\ee
In the rest of the article, we will drop the bar sign over the electrostatic potential in order to simplify the notation. The variational equations for the trial functions follow from the relations $\delta\Omega_v/\delta\phi_0(\br)=0$, $\delta\Omega_v/\delta\psi_0(\br)=0$, $\partial\Omega_v/\partial\kappa_c=0$, and $\partial\Omega_v/\partial\kappa_y=0$. The minimization yields
\begin{widetext}
\bea\label{eqvarEX1}
&&\Delta\phi_0(z)-\kappa^2_{DH}e^{-V_w(z)-V_c(z)-V_y(z)+\psi_b-\psi_0(z)}\sinh\phi_0(z)=-4\pi\ell_B q\sigma(z)\\
\label{eqvarEX2}
&&\Delta\psi_0(z)-b^2\psi_0+8\pi\ell_y\rho_be^{-V_w(z)-V_c(z)-V_y(z))+\psi_b-\psi_0(z)}\cosh\phi_0(z)=0
\eea
\bea\label{eqvarSC1}
&&\kappa_c^2=\kappa^2_{DH}\lan e^{-V_c(z)-V_y(z)+\psi_b-\psi_0(z)}\cosh\phi_0(z)
\frac{\partial V_c}{\partial\kappa_c}\ran_p\lan\frac{\partial V_c}{\partial\kappa_c}\ran_p^{-1}\\
\label{eqvarSC2}
&&\kappa_y^2=b^2+8\pi\ell_y\rho_b\lan e^{-V_c(z)-V_y(z)+\psi_b-\psi_0(z)}\cosh\phi_0(z)
\frac{\partial V_y}{\partial\kappa_y}\ran_p\lan\frac{\partial V_y}{\partial\kappa_y}\ran_p^{-1}
\eea
\end{widetext}
We will call the closure Eqs.~(\ref{eqvarEX1})-(\ref{eqvarSC2}) the general variational scheme (GVS). In the above relations, we defined the pore average as $\lan\cdot\ran_p=\int_0^d\mathrm{d}z\cdot/d$.  We also introduce the partition coefficient of coions $k_-$ and counterions $k_+$, that is, their pore averaged density renormalized with their bulk density as $k_\pm=\lan\rho_\pm(z)/\rho_b\ran_p$. For neutral pores where $\phi_0(z)=0$, one naturally gets $k_-=k_+=k$. The numerical implementation of the self-consistent relations~(\ref{eqvarEX1})-(\ref{eqvarSC2}) is briefly explained in Appendix~\ref{numsol}.

Eq.~(\ref{eqvarEX1}) is a modified PB equation that take into account pore-modified correlation effects associated with electrostatic and core interactions. Eq.~(\ref{eqvarEX2}) yields the local value of the external Yukawa potential that embodies the wetting effect issued from particle collisions. Finally, Eqs.~(\ref{eqvarSC1}) and~(\ref{eqvarSC2}) takes into account the modification of the screening of Yukawa and Coulomb interactions in the slit pore. The equations~(\ref{eqvarEX2}),~(\ref{eqvarSC1}) and~(\ref{eqvarSC2}) were solved in Ref.~\cite{jstat} for the case of neutral pores, where the external electrostatic potential $\phi_0(z)$ vanishes, in order to understand the role of the ion size in the mechanism of dielectric exclusion. In this article, we will first extend this study to the case of charged pores by solving the full set of equations~(\ref{eqvarEX1})-(\ref{eqvarSC2}) and also use the numerical solution for the variational functions $\kappa_y$, $\kappa_c$, $\psi_0(z)$ and $\phi_0(z)$ to compute the pressure between the plates, whose derivation is explained below.

\subsubsection{Evaluation of interplate pressure from GVS}

The derivation of the interplate pressure requires an explicit evaluation of the Grand potential in Eq.~(\ref{grandPot}). The details are explained in Appendix~\ref{appendixChr}. The result reads
\bea\label{grandPot2}
&&\frac{\Omega_v}{S}=\frac{1}{S}\left(\Delta\Omega_{0\phi}+\Delta\Omega_{0\psi}\right)\\
&&+\int_0^d\mathrm{d}z\left\{-\frac{\left[\nabla\phi_0(z)\right]^2}{8\pi\ell_B(z)q^2}+\frac{\sigma(z)}{q}\phi_0(z)\right\}\nonumber\\
&&-\int_0^d\frac{\mathrm{d}z}{8\pi\ell_y}\left\{\left[\nabla\psi_0(z)\right]^2+b^2\psi_0^2(z)\right\}
-\frac{b\left[\psi_0^2(0)+\psi_0^2(d)\right]}{8\pi\ell_y}\nonumber\\
&&-\int_0^d\mathrm{d}z\left[\rho_+(z)+\rho_-(z)\right],\nonumber
\eea
where the linear part of the corrections to the MF theory writes
\bea
&&\frac{1}{S}\Delta\Omega_{0\psi}=\frac{d}{24\pi}(\kappa_y-b)(\kappa_y^2+\kappa_y b-2b^2)\\
&&+\frac{b^2}{8\pi}\ln\frac{4b\kappa_y}{(b+\kappa_y)^2}+\int_0^\infty\frac{\mathrm{d}kk}{4\pi}\ln\left(1-\Delta_y^2e^{-2\rho_yd}\right)\nonumber\\
&&-\frac{\kappa_y^2-b^2}{8\pi}\int_0^\infty\frac{\mathrm{d}kk\Delta_y}{\rho_y^2}\frac{\Delta_y^2+2d\rho_y\Delta_y-1}
{1-\Delta_y^2e^{-2\rho_yd}}e^{-2\rho_yd}\nonumber
\eea
for core interactions and
\bea
&&\frac{1}{S}\Delta\Omega_{0\phi}=\frac{\kappa_c^3d}{24\pi}+\frac{\Delta_0\kappa_c^2}{16\pi}
+\int_0^\infty\frac{\mathrm{d}kk}{4\pi}\ln\left(1-\Delta_c^2e^{-2\rho_cd}\right)\nonumber\\
&&-\frac{\kappa_c^2}{8\pi}\int_0^\infty\frac{\mathrm{d}kk\Delta_c}{\rho_c^2}\frac{\Delta_c^2+2d\rho_c\Delta_c-1}
{1-\Delta_c^2e^{-2\rho_cd}}e^{-2\rho_cd}
\eea
for electrostatic interactions. We note that the functions $\Delta_0$, $\Delta_c$, $\Delta_y$, $\rho_c$ and $\rho_y$ are reported in Appendix~\ref{appendixCoef}.

The interplate pressure is defined as the derivative of the Grand potential Eq.~(\ref{grandPot2}) with respect to $d$ minus the bulk pressure,
\be
\beta\Pi=-\frac{1}{S}\frac{\partial\Delta\Omega_v}{\partial d}-\beta\Pi_b,
\ee
where the bulk pressure follows from Eq.~(\ref{grandPot2}) in the limit $V=Sd\to\infty$ as
\be\label{BulkPr}
\beta\Pi_b=2\rho_b-\frac{\kappa_{DH}^3}{24\pi}+\psi_b\rho_b-\frac{1}{24\pi}(\kappa_{yb}-b)(\kappa_{yb}^2+\kappa_{yb}b-2b^2).
\ee
The first two terms on the rhs. of this equation respectively correspond to the well-known entropic and electrostatic contributions to the bulk pressure. The third term takes into account the excluded volume of ions at the MF level, and the last term results from correlation effects associated with the repulsive Yukawa interactions. By taking into account the fact that the variation of the Grand potential Eq.~(\ref{grandPot2}) with respect to $\kappa_c$, $\kappa_y$, $\phi_0$ and $\psi_0$ vanishes, the interplate pressure takes the form
\bea\label{PrCor}
\Pi&=&-\frac{1}{S}\frac{\partial}{\partial d}(\Delta\Omega_{0\psi}+\Delta\Omega_{0\phi})-2\pi\ell_B\sigma_s^2-\psi_b\rho_b\nonumber\\
&&+\frac{\kappa_{DH}^3}{24\pi}+
\frac{1}{24\pi}(\kappa_{yb}-b)(\kappa_{yb}^2+\kappa_{yb}b-2b^2)\nonumber\\
&&+\rho_+(d)+\rho_-(d)-2\rho_b\nonumber\\
&&-\int_0^d\mathrm{d}z\left[\rho_+(z)+\rho_-(z)\right]\left(\frac{\partial V_c}{\partial d}+\frac{\partial V_y}{\partial d}\right)
\eea

The relation~(\ref{PrCor}) indicates that ionic penetration into the pore makes a repulsive contribution to the interplate pressure at three different levels. The first contribution is the screening of attractive vdW forces contained in the term $\Delta\Omega_\phi$. The second repulsive effect corresponding to the third line of the equation results from the translational entropy of ions within the pore. Finally, the last effect contained in the fourth line is an energetic contribution. This integral term is a pore-averaged force acting on the plates, and weighted by the ion density. The force in question is related to the energetic cost (associated with repulsive solvation and image charge interactions) to bring an ion from the reservoir into the pore. In Sec. III, Eq.~(\ref{PrCor}) will be investigated in the case of neutral and charged pores.

\subsubsection{Restricted variational scheme}

We will also propose in this article a restricted variational approach that simplifies the closure equations~(\ref{eqvarEX1})-~(\ref{eqvarSC2}). The calculation scheme is inspired by the variational Donnan approximation developed in Refs.~\cite{PRE,PRL,JCP}. Within this approximation, by taking into account the weak variations of the electrostatic potential in small pores and in the presence of weak surface charges, one replaces the spatially varying potential $\phi_0(z)$ with a constant trial potential $\phi_D$, whose value follow from the minimization of the Grand potential. It was shown in Ref.~\cite{PRE} that in the case of an electrolyte without core interactions, the approximation agrees quite well with the more general variational equations for pore averaged quantities such as partition coefficients. This is the approximation that we will adopt for the surface charge induced electrostatic interactions. However, the same approximation cannot work for Yukawa interactions, because it is impossible to satisfy the continuity of the derivative of a constant $\psi_0$ at the surface. Indeed, an inspection of Eq.~(\ref{eqvarEX2}) suggests that an effective external Yukawa potential $\psi_D$ should be rather introduced around the zero density solution of this equation, that is, the full potential should be of the form $\psi_0(z) =\psi_D+c\cosh\left[b(d/2-z)\right]$, where $c$ is a constant that should be fixed by the boundary condition Eq.~(\ref{boundpsi2}). Finally, the potential takes the form
\be\label{YukDon}
\psi_0(z)=\psi_D\left\{1-e^{-bd/2}\cosh\left[b(d/2-z)\right]\right\}.
\ee
By injecting the trial potentials $\phi_D$ and $\psi_0(z)$ into Eq.~(\ref{grandPot}), the Grand potential becomes
\bea\label{grandPotDON}
\frac{\Omega_v}{S}&=&\frac{1}{S}\left(\Omega_{0\phi}+\Omega_{0\psi}\right)\\
&&+\frac{2\sigma_s}{q}\phi_D+\frac{b\psi_D^2}{8\pi\ell_y}\left(1-bd-e^{-bd}\right)\nonumber\\
&&-\int_0^d\mathrm{d}z\left[\frac{\kappa_c^2}{8\pi\ell_B}v_0(\br,\br)
+\frac{\kappa_y^2-b^2}{8\pi\ell_y}w_0(\br,\br)\right]\nonumber\\
&&-2\rho_b\cosh\phi_D\int_0^d\mathrm{d}ze^{-V_c(z)-V_y(z)+\psi_b-\psi_0(z)}.\nonumber
\eea
The first variational equation $\delta\Omega_v/\delta\phi_D=0$ yields
\be\label{DonEl}
2\Gamma\sinh\phi_D=-\gamma,
\ee
where we defined the ratio between the pore volume density of the fixed surface charge and the bulk ionic density as $\gamma=2|\sigma_s|/(qd\rho_b)$, and
\be
\Gamma=\lan e^{-V_c(z)-V_y(z)+\psi_b-\psi_0(z)}\ran_p.
\ee
We note that for neutral pores, one has $\Gamma=k$. Eq.~(\ref{DonEl}) is clearly an electroneutrality relation taking into account image charge and core interactions. By inverting this relation in order to express $\phi_D$ in terms of the model parameters and injecting the solution into the remaining variational equations $\delta\Omega_v/\delta\kappa_c=0$, $\delta\Omega_v/\delta\kappa_y=0$, and $\delta\Omega_v/\delta\psi_D=0$, one ends up with the following non-linear equations
\begin{widetext}
\bea\label{eqvarSC3}
&&\kappa_c^2=\kappa^2_{DH}\sqrt{1+\frac{\gamma^2}{4\Gamma^2}}\lan e^{-V_c(z)-V_y(z)+\psi_b-\psi_0(z)}
\frac{\partial V_c}{\partial\kappa_c}\ran_p\lan\frac{\partial V_c}{\partial\kappa_c}\ran_p^{-1}\\
\label{eqvarSC4}
&&\kappa_y^2=b^2+8\pi\ell_y\rho_b\sqrt{1+\frac{\gamma^2}{4\Gamma^2}}\lan e^{-V_c(z)-V_y(z)+\psi_b-\psi_0(z)}
\frac{\partial V_y}{\partial\kappa_y}\ran_p\lan\frac{\partial V_y}{\partial\kappa_y}\ran_p^{-1}\\
\label{eqvarEX3}
&&\psi_D=\frac{bd\psi_b}{-1+bd+e^{-bd}}\sqrt{1+\frac{\gamma^2}{4\Gamma^2}}\lan e^{-V_c(z)-V_y(z)+\psi_b-\psi_0(z)}F(z)\ran_p.
\eea
\end{widetext}
We introduced in Eq.~(\ref{eqvarEX3}) the auxiliary function $F(z)=\psi_0(z)/\psi_D=1-e^{-bd/2}\cosh\left[b(d/2-z)\right]$. We will call the calculation scheme of Eqs.~(\ref{eqvarSC3})-(\ref{eqvarEX3}) the restricted variational scheme (RVS). It can be analytically shown that in the limit $d\to\infty$, these equations are reduced to $\kappa_c=\kappa_{DH}$, $\kappa_y=\kappa_{yb}$ and $\psi_D=\psi_b$. The meaning of the potential $\psi_D$ can be understood by injecting Eq.~(\ref{YukDon}) into Eq.~(\ref{eqvarEX2}) and integrating once from $z=0$ to $z=d$. One immediately gets $\psi_D=\psi_b(k_++k_-)/2$. This equality shows that $\psi_D$ accounts for the modification of the bulk Yukawa potential $\psi_b$ by pore modified correlation effects. Strictly speaking, the value of $\psi_D$ given by this equality differs from the one obtained from Eq.~(\ref{eqvarEX3}). However, we verified that both values coincide with a very good accuracy for the range of parameters considered in this article (data not shown). Furthermore, the partition coefficients can be rewritten as
\be\label{partcoDON}
k_\pm=\frac{1}{2}\left(\sqrt{\gamma^2+4\Gamma^2}\pm\gamma\right).
\ee

\subsubsection{Evaluation of interplate pressure from RVS}

The interplate pressure within RVS is obtained by taking the derivative of the Grand potential Eq.~(\ref{grandPotDON}) with respect to $d$. By taking into account the fact that $\Omega_v$ is optimum with respect to $\kappa_c$, $\kappa_y$, $\phi_D$, and $\psi_D$, the pressure becomes
\bea\label{PrCorDON}
\Pi&=&-\frac{1}{S}\frac{\partial}{\partial d}(\Delta_{0\phi}+\Omega_{0\psi})+\frac{b^2\psi_D^2}{8\pi\ell_y}\left(1-e^{-bd}\right)-\psi_b\rho_b\nonumber\\
&&+\frac{\kappa_{DH}^3}{24\pi}+\frac{1}{24\pi}(\kappa_{yb}-b)(\kappa_{yb}^2+\kappa_{yb}b-2b^2)\\
&&+\rho_+(d)+\rho_-(d)-2\rho_b\nonumber\\
&&-\int_0^d\mathrm{d}z\left[\rho_+(z)+\rho_-(z)\right]\left(\frac{\partial V_c}{\partial d}+\frac{\partial V_y}{\partial d}+\frac{\partial\psi_0}{\partial d}\right).\nonumber
\eea

The importance of RVS Eqs.~(\ref{eqvarSC3})-(\ref{eqvarEX3}) is threefold. First of all, they significantly simplify the numerical investigation of Eqs.~(\ref{eqvarEX1})-~(\ref{eqvarSC2}). We note that self consistent equations derived with drastic approximations and similar in form to Eq.~(\ref{eqvarSC1}) have been used in theoretical as well as experimental nanofiltration studies~\cite{yarosch,YarII,Szymczyk}. Hence, the relations~(\ref{eqvarSC3})-(\ref{eqvarEX3}) present themselves as a practical way to account for excluded volume effects in these studies. Secondly, they would allow to investigate these effects in more complicated geometries, such as spherical colloid systems or cylindrical ion channels, where the integration of Eqs.~(\ref{eqvarEX1})-(\ref{eqvarEX2}) becomes quite involved~\cite{JCP}. Finally, the proposed scheme is equally promising for understanding in pore geometries further ion specific effects, such as hydration interactions, whose consideration would add to the closure relations~(\ref{eqvarEX1})-(\ref{eqvarSC2}) further external fields associated with these forces~\cite{PRLOrl2}. In the next section, the prediction of RVS for ion densities and interplate pressure will be compared with the result of GVS equations.

\section{Numerical results}

We investigate in this section excluded volume effects induced by core collisions on the interaction between two planar walls (see Fig.~\ref{sketch}).  The MF level computation of the interplate pressure $\Pi$ consists in evaluating Eq.~(\ref{MFPr2}) with the numerical solution of Eqs.~(\ref{MFphi}) and ~(\ref{MFpsi}). The calculation of $\Pi$ within GVS given by Eq.~(\ref{PrCor}) is carried out after an iterative solution of  Eqs.~(\ref{eqvarEX1})-(\ref{eqvarSC2}) that yields $\kappa_c$, $\kappa_y$, $\phi_0(z)$, and $\psi_0(z)$ (see Appendix~\ref{numsol} for details). We will also compare the results obtained from these equations with the RVS of Eqs.~(\ref{eqvarSC3})-(\ref{eqvarEX3}).

\subsection{Neutral particles with core interactions}
\begin{figure}
(a)\includegraphics[width=1.1\linewidth]{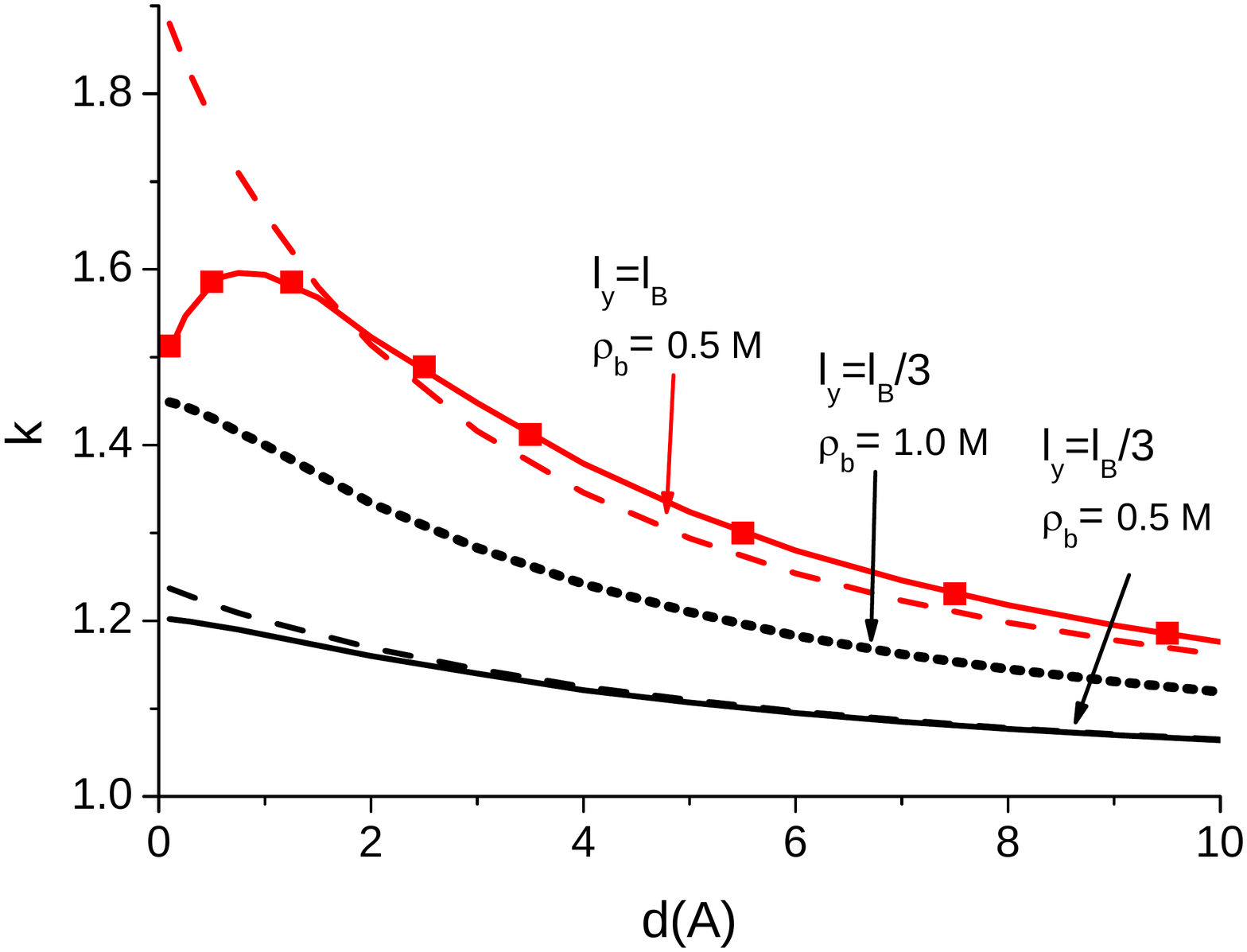}
(b)\includegraphics[width=1.1\linewidth]{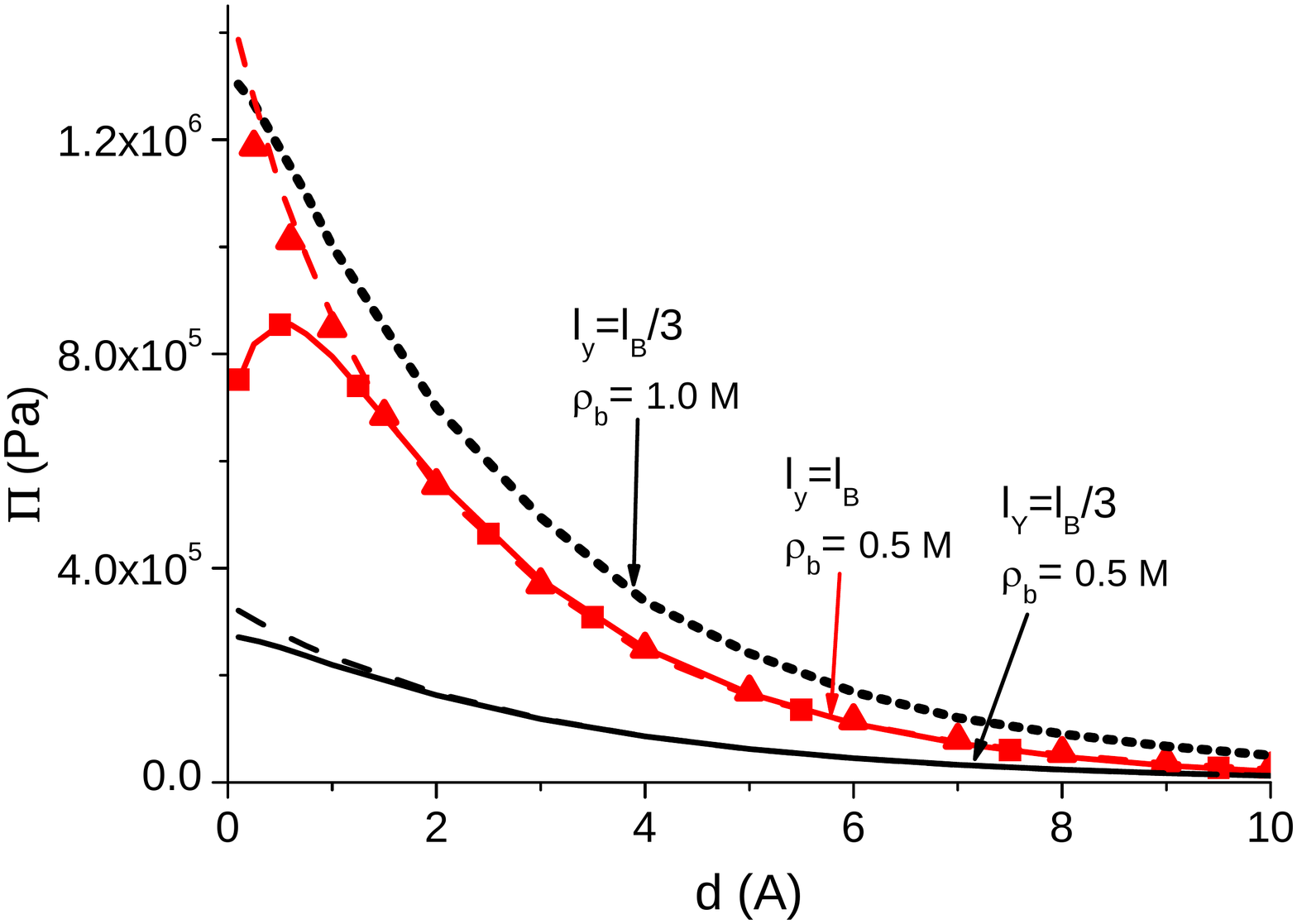}
\caption{(Color online) (a) Partition coefficients and (b) interplate pressure for neutral particles in the slit pore against the pore size for two values of $\rho_b$ and $\ell_y$. Solid and dotted lines are obtained from GVS, squares show the result of RVS, dashed curves are the MF results, and triangles in (b) mark the analytical expression~(\ref{lincore}).}
\label{PrNt}
\end{figure}

We illustrate in Fig.~\ref{PrNt} the partition coefficient of neutral Yukawa particles ($q=0$ and $\kappa_c=0$) and the interplate pressure mediated by core collisions between them against the pore size $d$ for $b=2/\ell_B$ and two values of $\ell_y$, $\rho_b$, and a vanishing surface charge. In order to evaluate exclusively the contribution from excluded volume effects, we consider the case $\e_m=\e_w$ where the unscreened vdW contribution $\Pi_{vdW}(\kappa_c=0)=-Li_3(\Delta_0^2)/(8\pi d^3)$ vanishes. First of all, Fig.~\ref{PrNt}.(a) shows that the pore density of neutral Yukawa particles exceeds their reservoir density. In other words, there exists a net particle adsorption into the pore. Repulsive core interactions within the bulk reservoir that drive the Yukawa particles into the confined pore medium are responsible for this effect~\cite{jstat}. Moreover, we see that according to the MF calculation, due to an intensification of particle packing induced by core collisions, this adsorption effect is amplified with decreasing $d$, or increasing $\ell_y$. Although the variational estimation of pore averaged particle density within GVS agrees with the MF one for intermediate to large values of $d$, the MF result deviates from the variational prediction below a characteristic pore size where  $k$ becomes non-monotonical. Indeed, this deviation originates from repulsive solvation forces (similar to the ones observed for purely electrostatic systems~\cite{Bratko1,Bratko2}), embodied in the potential $V_y(z)$ of Eq.~(\ref{PMFslit2}), that come into play exclusively at the variational level. It was shown in Ref.~\cite{jstat} that at small pore sizes, the magnitude of these forces may grow faster than core collision effects that drives particles into the pore (the term $\psi_b-\psi(z)$) with decreasing $d$. As a result, while decreasing $d$, the pore averaged particle density reaches a maximum and starts to decrease.

A comparison of Figs.~\ref{PrNt}.(a) and (b) shows that for the neutral Yukawa liquid, the behaviour of the interplate pressure is mainly dictated by particle packing into the pore. First of all, the particle excess within the slit pore yields a positive pressure between the walls. Then, one sees that at the MF level, a reduction of the pore size or an increase of $\ell_y$ and $\rho_b$ that amplifies particle adsorption leads to a monotonous increase of the interplate pressure. Moreover, while decreasing the pore size, at small values of $d$ where the deviation of the MF result from the variational one becomes noticeable, the variational pressure first reaches a repulsive peak and then starts decreasing with $d$. We also note that the analytical solution of the MF pressure Eq.~(\ref{lincore}) reported in Fig.~\ref{PrNt}.(b) shows a good agreement with the exact MF expression Eq.~(\ref{MFPr2}). Furthermore, it is shown that the results obtained with GVS and RVS show a very good agreement. To conclude, we plot in Fig.~\ref{PrNt} the partition coefficient and the interplate pressure for a larger bulk density $\rho_b=1$ M and a weaker Yukawa coupling $\ell_y=\ell_B/3$ in order to show that an increase of $\rho_b$ at fixed $\ell_y$ is equivalent to an increase of $\ell_y$ at fixed $\rho_b$, that is, both enhance the strength of excluded volume effects.

In the case where the particles posses a finite charge, electrostatic interactions that come into play will compete with the excluded volume effects discussed so far. This competition will be thoroughly investigated in the following parts.

\subsection{Interplay between electrostatic forces and core interactions}

\begin{figure}
(a)\includegraphics[width=1.1\linewidth]{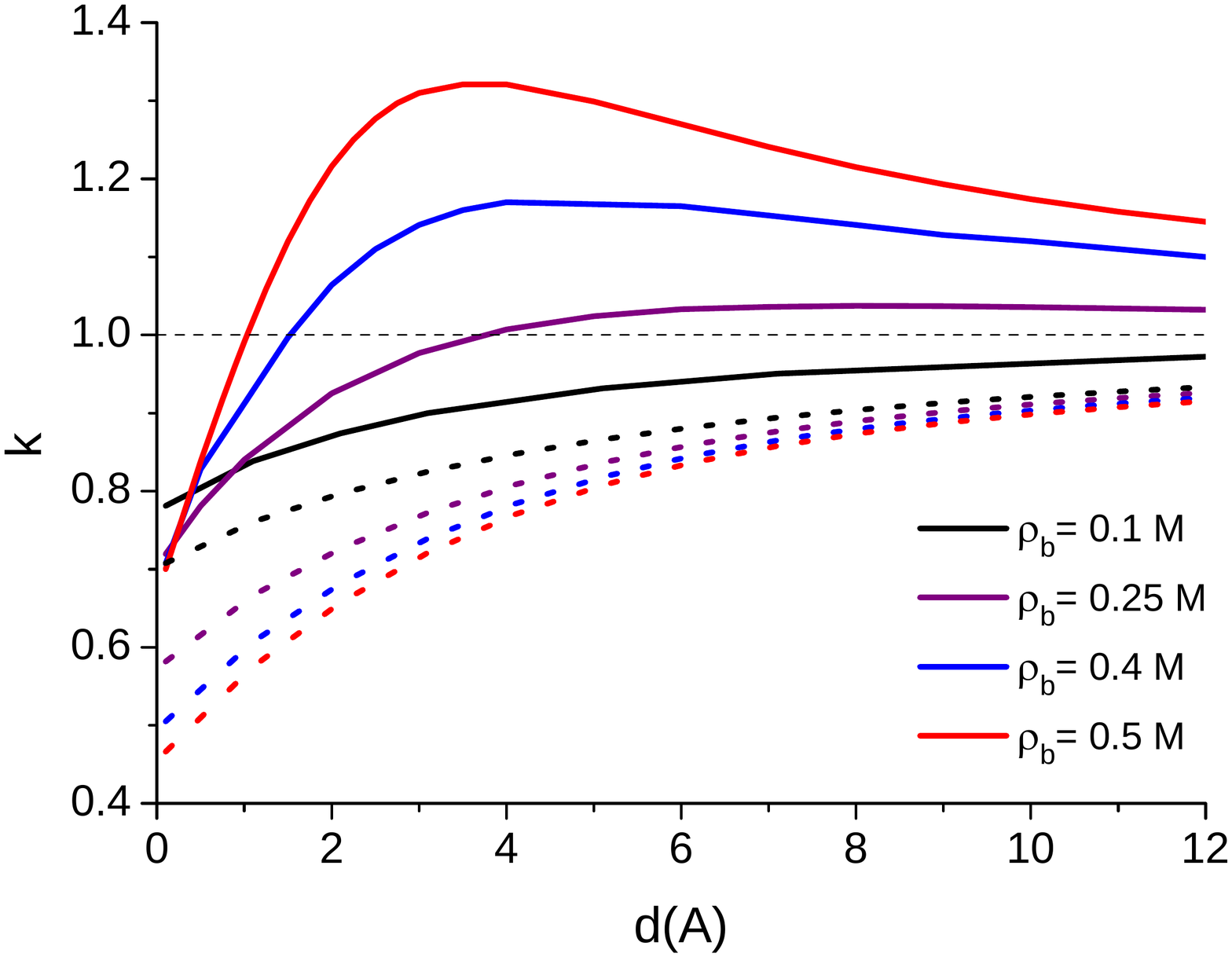}
(b)\includegraphics[width=1.1\linewidth]{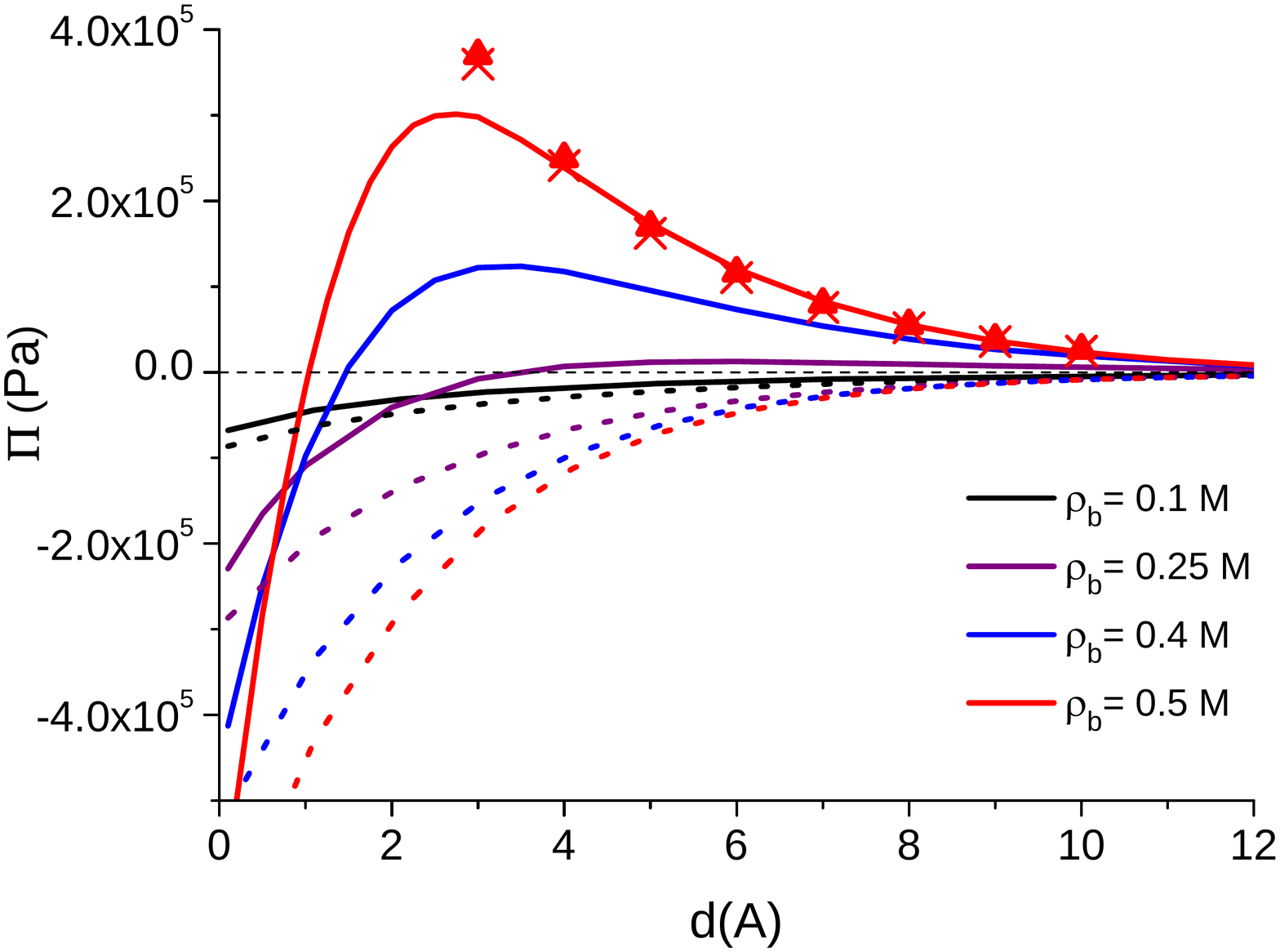}
\caption{(Color online) (a) Partition coefficients and (b) interplate pressure against the pore size for charged particles with $\ell_y=0$ (dotted lines), $\ell_y=\ell_B$ and $b=2/\ell_B$ (solid lines) in the case $\e_m=\e_w$. Solid and dotted curves are obtained from GVS, and the crosses and triangles are respectively the non-linear and linear MF predictions of Eqs.~(\ref{MFPr1}) and~(\ref{lincore}).}
\label{prch1}
\end{figure}

The results presented in this part are derived in the case of a vanishing surface charge ($\sigma_s=0$) where the electrostatic potential $\phi_0(z)$ vanishes. The variational results are obtained from GVS and they will be compared with RVS in all cases. Fig.~\ref{prch1} displays the ionic partition coefficient and the net interplate pressure against the pore size for Coulomb and charged Yukawa particles of valency $q=1$, with $\ell_y=\ell_B$, $b=2/\ell_B$, and several values of $\rho_b$ in the case $\e_m=\e_w$ where image forces vanish. As in the case of neutral Yukawa particles, one notices the direct correlation between $k$ and $\Pi$. For dilute electrolytes (e.g. the curve for $\rho_b=0.1$ M), one sees that there is a net particle exclusion from the pore (i.e. $k<1$) and the pore-averaged density of the Yukawa electrolyte is very close to that of the Coulomb liquid. This weak depletion effect is known to be due to electrostatic solvation forces associated with the distortion of the ionic cloud by the pore walls, and leads to a negative interplate pressure~\cite{hatlo,PRE}.

For the lowest bulk concentration considered in Fig.~\ref{prch1} (i.e. $\rho_b=0.1$ M), the increase of the pore size is accompanied with the weakening of solvation forces and an increase of the pore ionic density towards the bulk value. As seen in the top and bottom plots of Fig.~\ref{prch1}, the ionic penetration results in a monotonous decay of the negative pressure with increasing $d$. Furthermore, one notices that with increasing $\rho_b$, the deviation of $k$ and $\Pi$ from the purely Coulombic case ($\ell_y=0$) becomes more pronounced. Namely, the evolution of the partition coefficient and the pressure with the pore size becomes non-monotonic. In the small $d$ regime, ionic penetration into the pore is mainly driven by repulsive solvation forces, which leads to a net particle depletion and a negative pressure. In the large $d$ regime where excluded volume effects induced by core interactions begin to take over solvation forces, the pore density of particles exceeds their reservoir density. As the consequence of this ionic packing in the pore, the interaction between the pore walls changes its sign and becomes repulsive. With further increase of the pore size, $\lan\rho\ran$ reaches a maximum and starts decreasing towards $\lan\rho_b\ran$. In this large $d$ regime where core interactions play the main role in ionic penetration, $\Pi$ follows the same trend as $k$. Namely, $\Pi$ reaches a repulsive peak and begins to decrease with increasing $d$. We note that this interpolation of $k$ and $\Pi$ between a particle depletion limit with $\Pi<0$ and a particle excess regime with $\Pi>0$ was also observed in MC simulations of Refs.~\cite{Bratko1,Bratko2} for charged particles with HC interactions confined in slit pores without dielectric discontinuity. It is also shown in Fig.~\ref{prch1} that for $d>4$ {\AA}, the linear and non-linear MF predictions of Eqs.~(\ref{MFPr1}) and~(\ref{lincore}) for the interplate pressure agree well with the variational result, which enables us to identify this large pore size range $d>4$ {\AA} as the MF regime of the model. The failure of the MF theory for small pores can be simply explained by the fact that for neutral pores, electrostatic interactions are not taken into account at the MF level.
\begin{figure}
\includegraphics[width=1.1\linewidth]{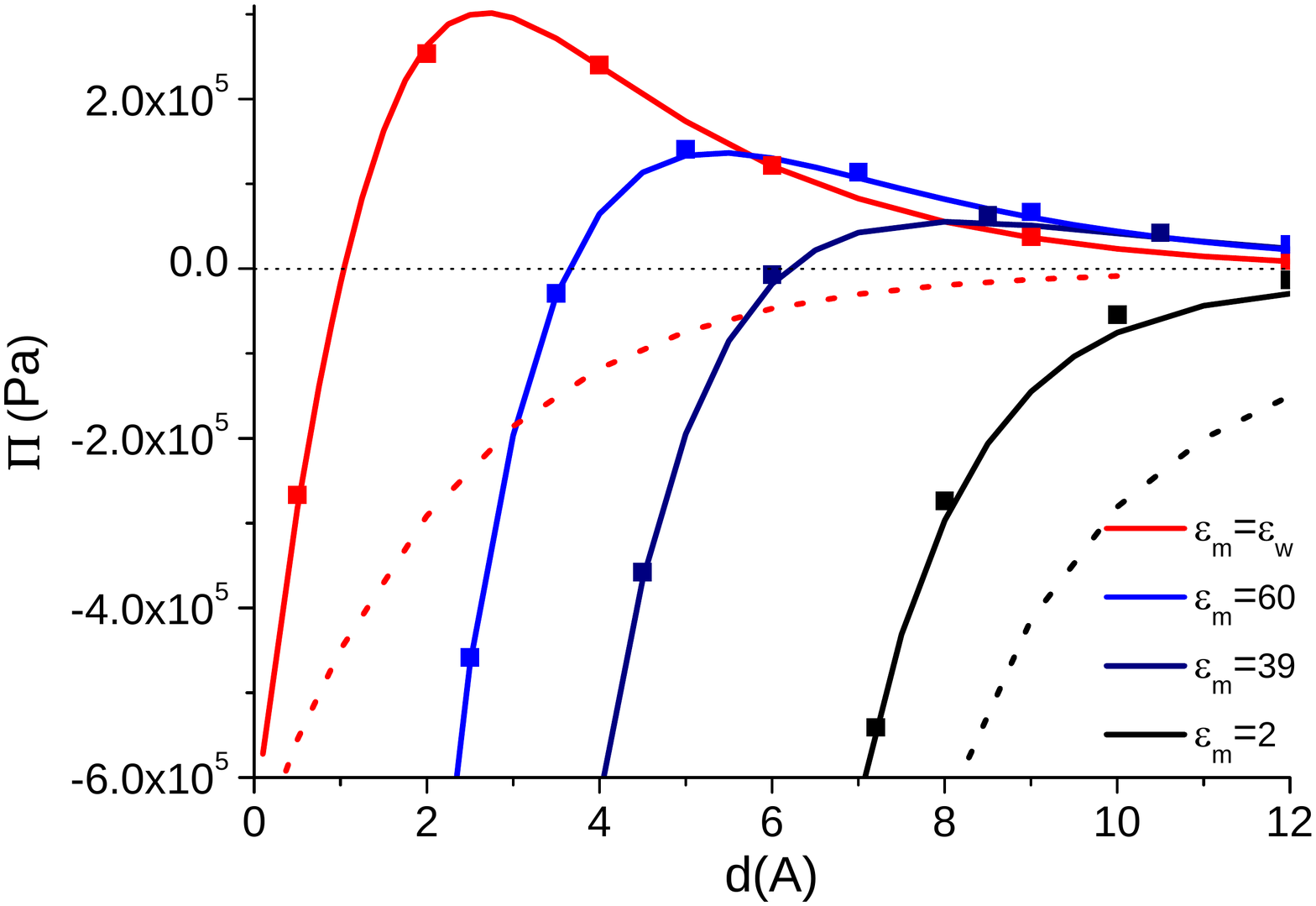}
\caption{(Color online) Interplate pressure for charged particles with $\ell_y=\ell_B$, $b=2/\ell_B$, and  $\rho_b=0.5$ M, for various $\e_m$. Solid and dotted curves that correspond respectively to the Yukawa and the Coulomb liquid are obtained from GVS. The squares show the prediction of RVS.}
\label{prchvsem}
\end{figure}

We finally note that the presence of core interactions reverses the behaviour of $k$ and $\Pi$ with respect to a change in $\rho_b$. Namely, for the Coulomb fluid (dotted lines in Fig.~\ref{prch1}.(a)), an increase of $\rho_b$ that amplifies repulsive solvation forces leads to a reduction of $k$ (up to molar concentrations where these forces start to be screened). The increase of the particle depletion yields in turn an enhanced attractive pressure (see Fig.~\ref{prch1}.(b)).  In the case of the Yukawa liquid, in addition to the amplification of solvation forces, the increase of $\rho_b$ also amplifies the particle packing in the pore. Consequently, for interplate separations within the range $d>1$ {\AA} where the latter effect dominates the former one, the increase of $\rho_b$ at fixed $d$ from a small value makes the initially attractive pressure gradually more repulsive and pushes the equilibrium point where the pressure vanishes towards smaller interplate separations.
\begin{figure}
(a)\includegraphics[width=1.1\linewidth]{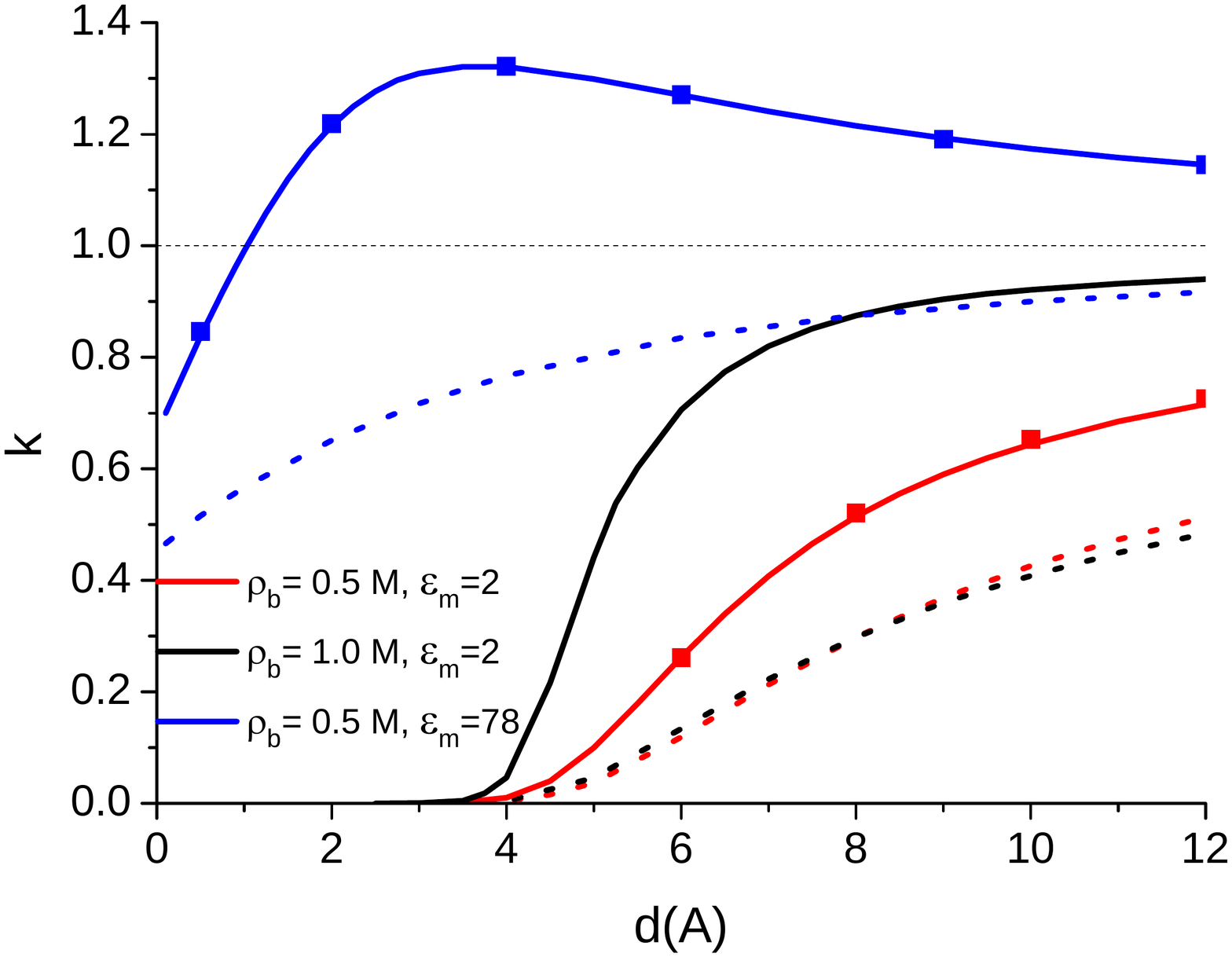}
(b)\includegraphics[width=1.1\linewidth]{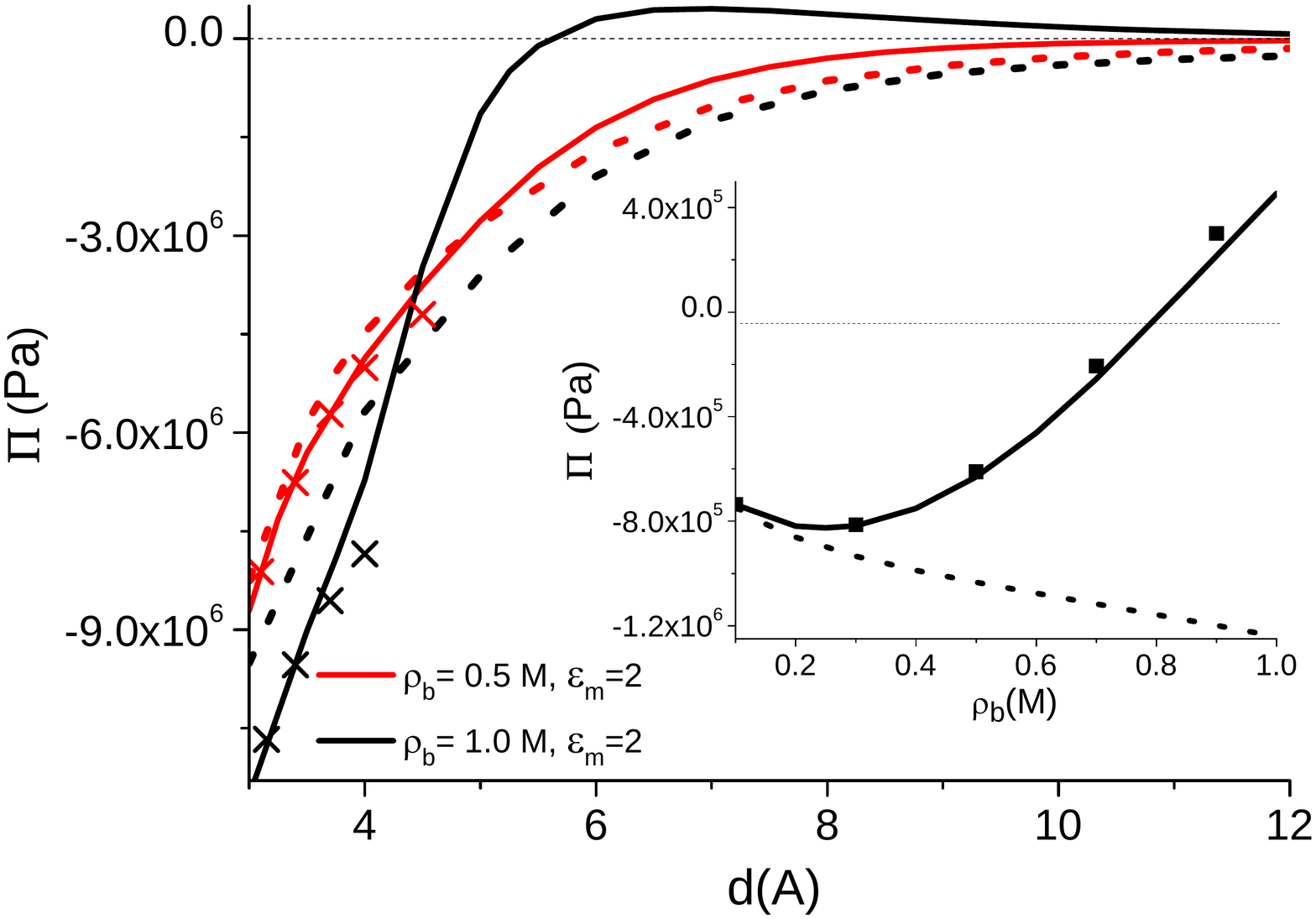}
\caption{(Color online) (a) Partition coefficients and (b) interplate pressure for charged particles with $\ell_y=0$ (dotted lines), $\ell_y=\ell_B$ (solid lines), and $b=2/\ell_B$. The inset in (b) displays $\Pi$ against $\rho_b$ for $d=7$ {\AA} and $\e_m=2$. Solid curves are obtained from GVS, crosses mark the dilute pore regime reported in the text, and the squares in (a) and in the inset of (b) display the prediction of RVS.}
\label{prch2}
\end{figure}

Figs.~\ref{prchvsem} and~\ref{prch2}.(a) display, respectively, the evolution of the interplate pressure and the partition coefficient of the Coulomb and the Yukawa liquids for $\ell_y=\ell_B$, $b=2/\ell_B$, $\rho_b=0.5$ M and several values of the matrix permittivity from $\e_m=2$ to $\e_m=\e_w$. A small value of $\e_m$ is known to yield a strong van der Waals energy and a pronounced ionic depletion~\cite{hatlo,PRE}. It is also known that both effects bring an attractive contribution to the pressure. As seen in these figures, the decrease of $\e_m$ at fixed $d$ has the effect of reducing the intensity of the particle adsorption into the pore induced by core collisions within the bulk, thereby shifting the equilibrium distance towards larger interplate separations. This delicate balance between excluded volume effects and image forces clearly indicates the former as an important ingredient to understand the stability of macromolecules in electrolyte solutions.

In order to better understand the role played by core collisions in the interaction between the walls of a dielectrically heterogeneous pore, we illustrate in Figs.~\ref{prch2}.(a) and (b) the partition coefficient and the interplate pressure of the Coulomb and Yukawa fluids against the pore size for $\e_m=2$ (the characteristic value for lipid bilayers) and two values of $\rho_b$. For small pore sizes where the dielectric exclusion mechanism considerably reduces the pore density of ions, the pressure Eq.~(\ref{PrCor}) reduces to $\Pi\simeq-\mathrm{Li}_3(\Delta_0^2)/(8\pi d^3)-\Pi_b$, where the outer osmotic pressure $\Pi_b$ is given by Eq.~(\ref{BulkPr}). This limiting law is reported in Fig.~\ref{prch2}.(b) (crosses). One first notices that in this small $d$ or dilute pore regime, for both types of ionic liquid, an increase of the bulk concentration that enhances the magnitude of the outer pressure $\Pi_b$ strengthens the attraction between the plates. Then, by comparing the curves corresponding to $\ell_y=0$ and $\ell_y=\ell_B$, one sees that in the same small pore limit, the presence of core interactions adds to the attractive force mediated by purely electrostatic interactions. The intensification of the attraction between the plates by core collisions originates from the Yukawa contribution to the bulk osmotic pressure $\Pi_b$.

An inspection of the solid curves for $\e_m=2$ in Fig.~\ref{prch2}.(a) shows that with increasing surface separation, the decay of repulsive image forces leads to a rapid increase of the ionic density within the pore. Moreover, by comparing at fixed $d$ the dotted and solid curves with concentrations $\rho_b=0.5$ and 1 M , we also note that as in the case without a dielectric discontinuity, the presence of core interactions makes $k$ a rapidly increasing function of $\rho_b$. Consequently, one sees in Fig.~\ref{prch2}.(b) that in the large surface separation regime where ionic packing within the pore becomes important ($d>5$ {\AA}), the increase of the bulk concentration of the Yukawa liquid makes the pressure less attractive. The non-monotonical behaviour of $\Pi$ with respect to a change in $\rho_b$ is also illustrated for $d=7$ {\AA} in the inset of Fig.~\ref{prch2}.(b). One also notices that for large enough $d$ and $\rho_b$ where the intensity of excluded volume effects become comparable with the magnitude of vdW and depletion forces responsible for the attractive part of the interplate pressure, $\Pi$ becomes purely repulsive, before reaching a repulsive peak and decaying with increasing $d$. We note that the transition of the pressure from an attractive to a repulsive regime with increasing interplate separation was equally observed within the AHNC method for concentrated electrolytes with HC interactions confined in dielectrically heterogeneous pores (see Fig.5 of Ref.\cite{kjal3}).
\begin{figure}
(a)\includegraphics[width=1.1\linewidth]{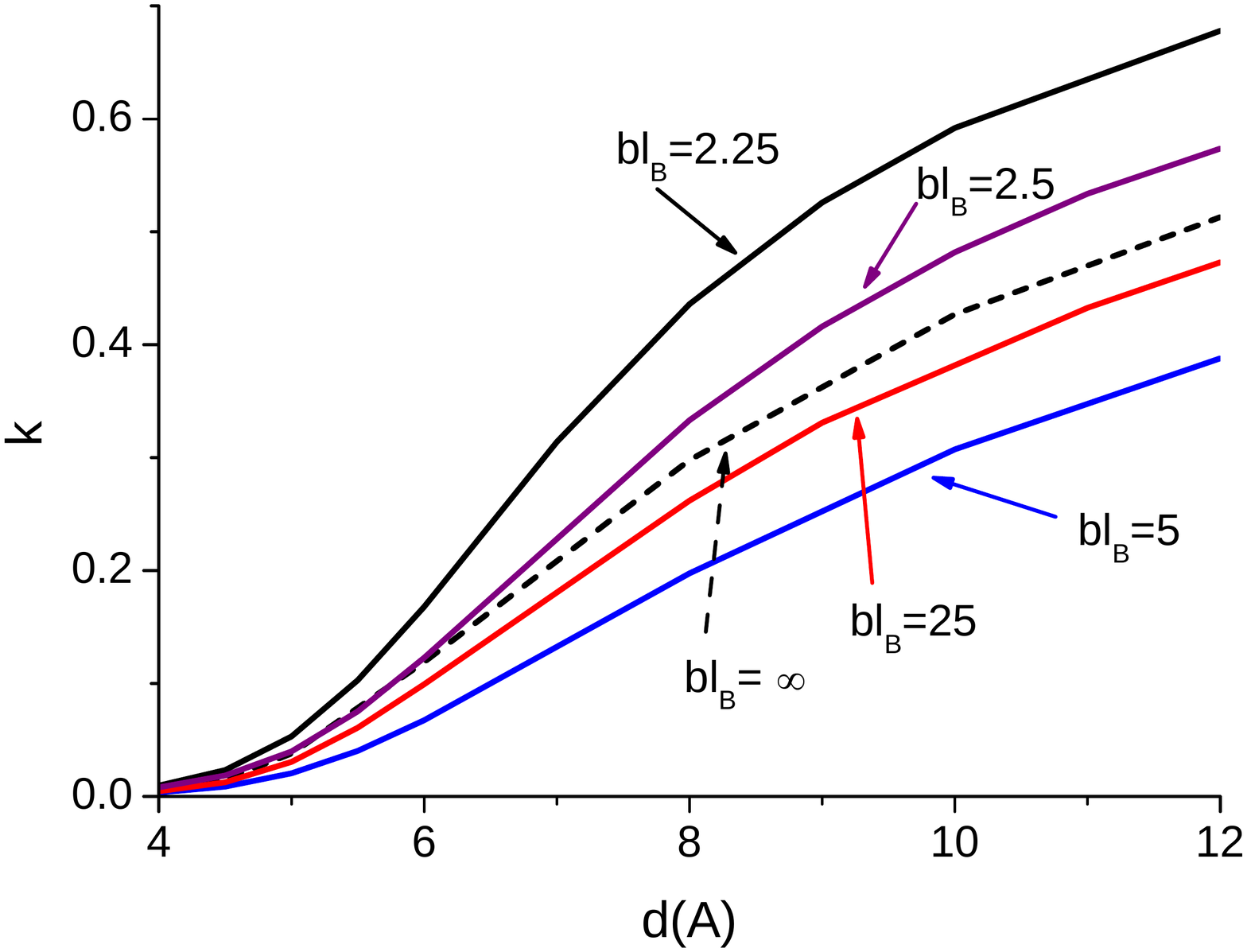}
(b)\includegraphics[width=1.1\linewidth]{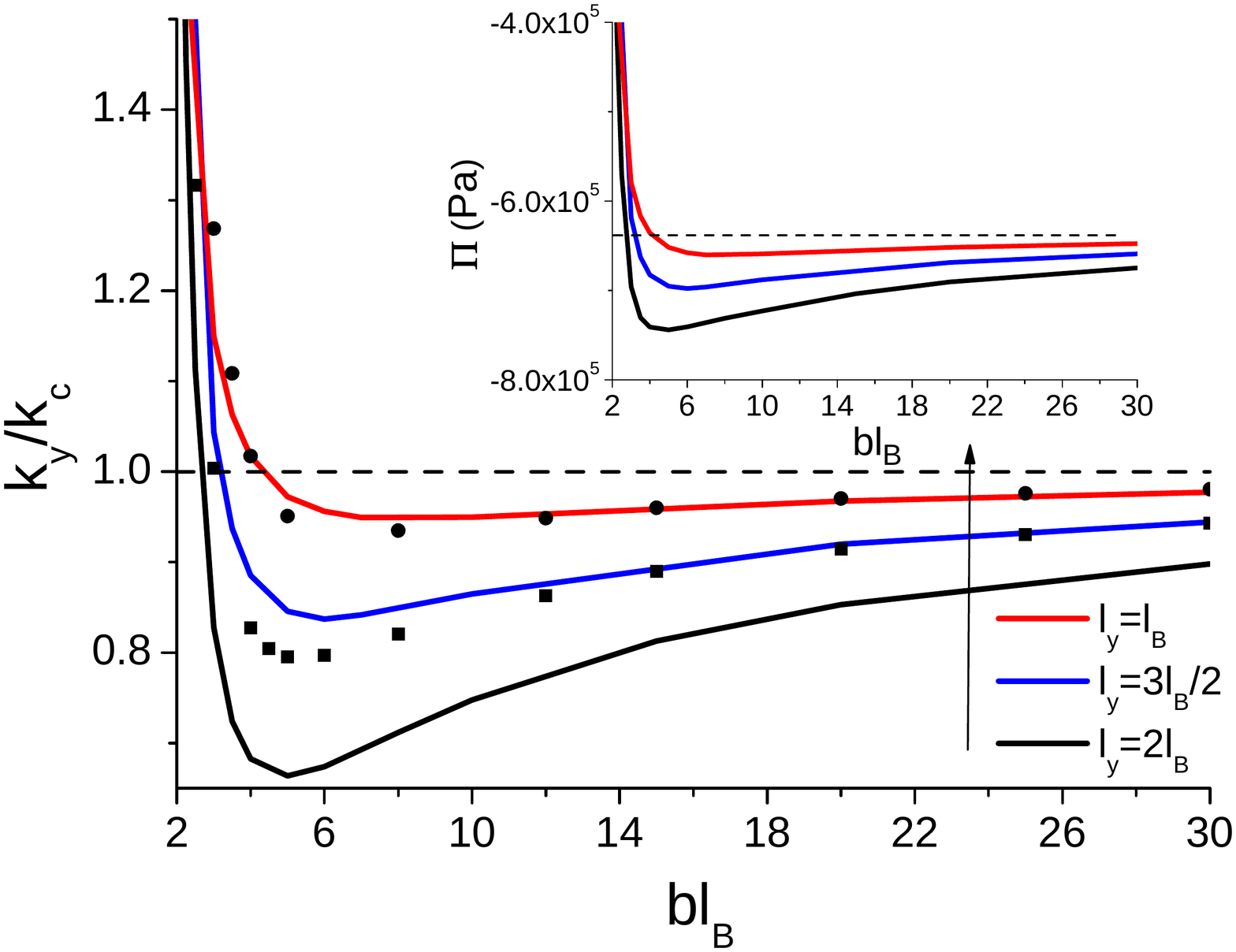}
\caption{(Color online) (a) Ionic partition coefficients versus the pore size at bulk density $\rho_b=0.5$ M. The membrane permittivity is $\e_m=2$. The dashed line corresponds to vanishing core interactions (i.e. Coulomb liquid), reached in the limit $\ell_y\to0$ or $b\ell_B\to\infty$. The solid lines show the case $\ell_y=2\ell_B$ for various $b$. (b) Ionic partition coefficients $k_y=k(\ell_y)$ renormalized with $k_c=k(\ell_y=0)$ against the screening length $b\ell_B$ at $d = 8$ {\AA}, $\rho_b=0.5$ M, $\e_m=2$ (solid lines), $\e_m=39$ (squares) and $\e_m=\e_w$ (circles). The inset displays the interplate pressure for the same model parameters and $\e_m=2$. The horizontal reference line marks the value of the pressure for $\ell_y=0$. All results are obtained from GVS.}
\label{prchvsb}
\end{figure}

We finally illustrate in Figs.~\ref{prchvsem} and~\ref{prch2} the prediction of the simple self consistent scheme Eqs.~(\ref{eqvarSC3})-(\ref{eqvarEX3}) for the partition coefficients Eq.~(\ref{partcoDON}) and the interplate pressure Eq.~(\ref{PrCorDON}) for a wide range of bulk ion density, pore size and membrane permittivity. It is seen that in all cases, the restricted scheme shows a very good agreement with the general one.

\subsection{Influence of the range of core interactions}

The results discussed in this part were all derived within GVS. In order to understand the role of the range of core interactions in the ionic exclusion mechanism and the interaction force between the pore walls that we still consider neutral, we first compare in Fig.~\ref{prchvsb}.(a) the partition coefficient of the Coulomb and Yukawa particles for $\ell_y=2\ell_B$ and four values of $b$. With an increase of $b$ at fixed $d$ that weakens core collision effects driving the ions from the bulk into the pore and also towards the interfaces, one would expect the partition coefficient of Yukawa charges $k_y=k(\ell_y=\ell_B)$  to monotonously decrease towards the partition coefficient of Coulomb charges $k_c=k(\ell_y=0)$. However, an inspection of Fig.~\ref{prchvsb}.(a) shows that the behaviour of $k_y$ deviates from this picture. First of all, with the increase of the bare screening parameter from $b\ell_B=2.25$ to $b\ell_B=5$ at the coupling parameter $\ell_y=2\ell_B$, $k_y$ drops below $k_c$. Then, if one continues to increase the screening parameter, $k_y$ changes its trend and begins to rise towards $k_c$, until it reaches in the limit $b\ell_B\to\infty$ the partition coefficient of the Coulomb liquid. As it will be shown below in detail, the effect of an additional ionic exclusion arising in the presence of strongly screened core interactions is mainly driven by the balance between core collisions within the reservoir that drive the particles into the pore, and solvation forces associated with core interactions that exclude them from the pore. We note that this effect is partly responsible for a higher rejection rate of finite size ions from dielectrically heterogeneous membranes~\cite{jstat}.
\begin{figure}
(a)\includegraphics[width=1.1\linewidth]{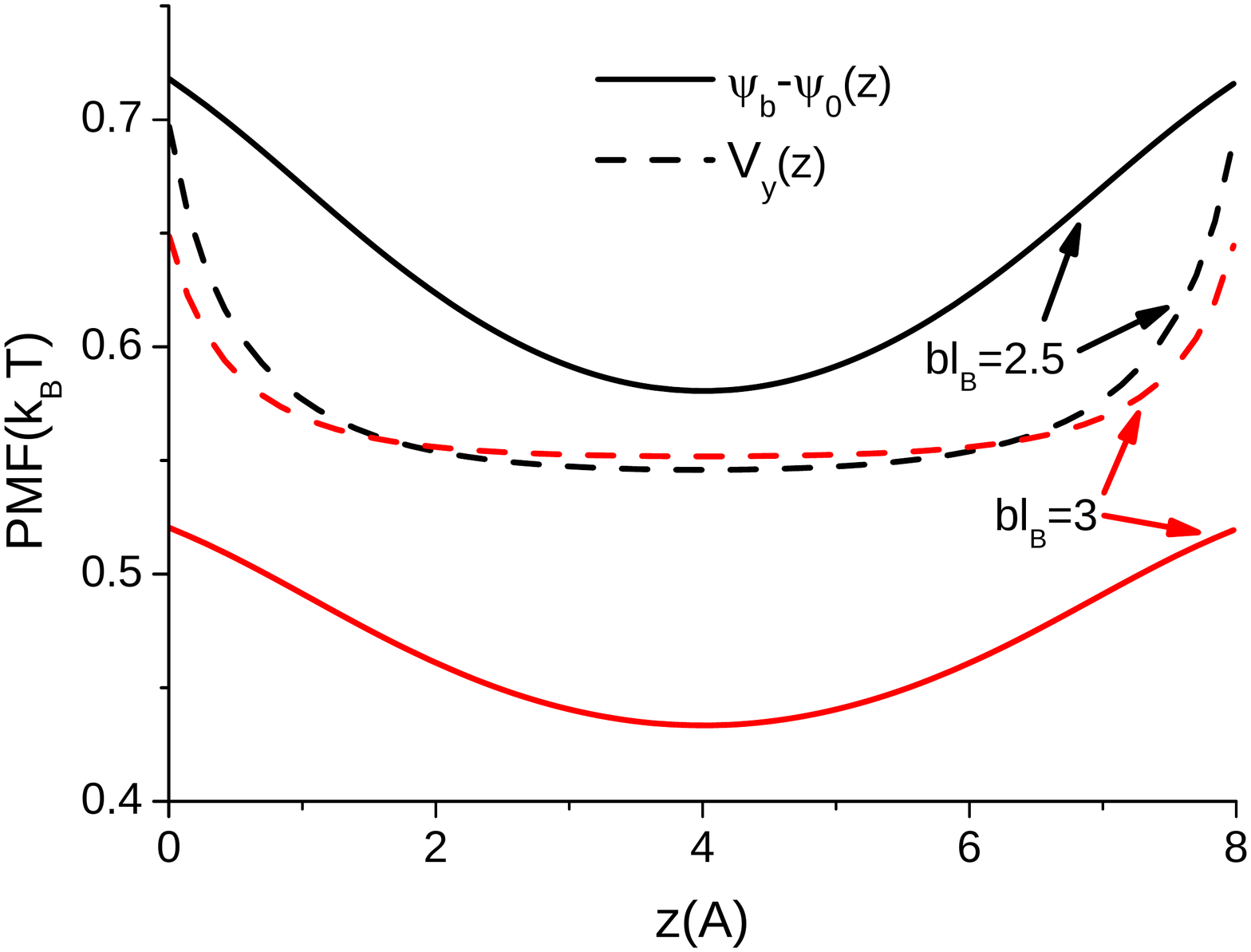}
(b)\includegraphics[width=1.1\linewidth]{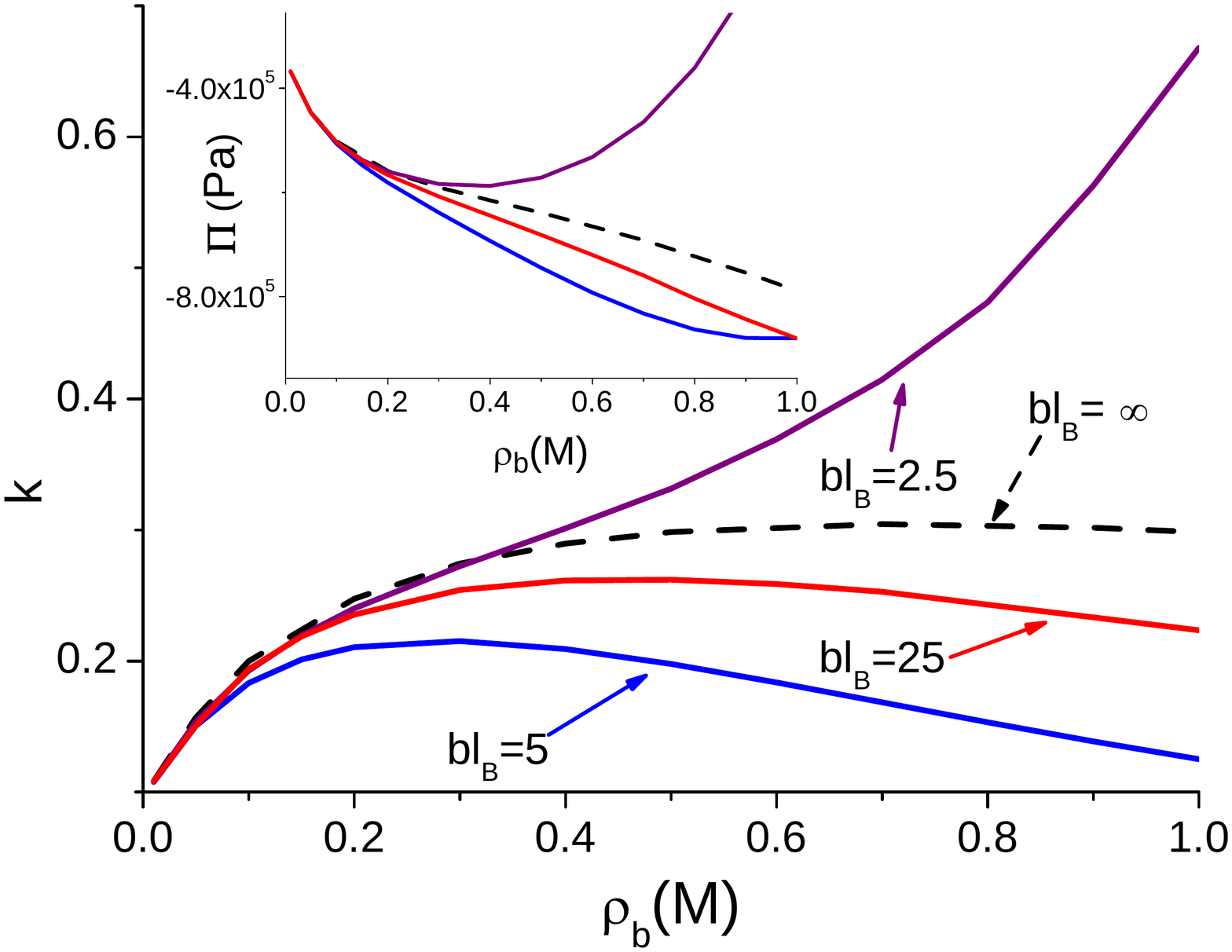}
\caption{(Color online)  (a) Yukawa potential profiles for $b\ell_B=2.5$ (black curves) and $b\ell_B=3.0$ (red curves), for the pore size $d=8$ {\AA}, the bulk concentration $\rho_b=0.5$ M, the coupling parameter $\ell_y=2\ell_B$ and the membrane permittivity $\e_m=2$. (b) Ionic partition coefficients against the bulk density at $d=8$ {\AA} and for the same model parameters as in Fig.~\ref{prchvsb}.(a). The inset illustrates the interplate pressure against $\rho_b$. All results are obtained from GVS.}
\label{prchvsb2}
\end{figure}

To progress further in the analysis of the range of Yukawa interactions, we plotted in Fig.~\ref{prchvsb}.(b) the reduced partition coefficient $k_y/k_c$ against $b\ell_B$ at fixed interplate separation ($d=8$ {\AA}), and for various values of $\ell_y$ and $\e_m$. The inset shows the interplate pressure for the same model parameters. The non-monotonic evolution of $k_y$ and $\Pi_y=\Pi(\ell_y>0)$ with $b$ is clearly illustrated in this figure. Namely, while increasing the screening parameter from $b\ell_B=2$, $k_y$  decreases and reaches $k_c$  at $b\ell_B=3$ to 5. This screening range corresponds to the regime of moderately screened core interactions discussed in Fig.~\ref{PrNt} to Fig.~\ref{prch2}, where core collision effects embodied in the potential term $\psi_0(z)-\psi_b$  of Eq.~(\ref{locden}) dominate solvation forces associated with these interactions, i.e. the potential $V_y(z)$ of the same equation. This balance leads to an ionic packing, that is $k_y>k_c$, which leads to $\Pi_y>\Pi_c=\Pi(\ell_y=0)$. With a further increase of $b$, one gets into the second regime of strongly screened core interactions where due to a faster decay of $\psi_0(z)-\psi_b$ with respect to $V_y(z)$, $k_y$  rapidly drops below $k_c$, until it reaches a minimum located at $b\ell_B\simeq5-7$. Moreover, it is seen in the inset of  Fig.~\ref{prchvsb}.(b) that $\Pi_y$ follows the same trend as $k_y$. We also show in Fig.~\ref{prchvsb2} the transition of the core interaction energy $V_y(z)-\psi_b+\psi_0(z)$ from a negative to a positive value with an increase of $b\ell_B$ from 2.5 to 3. One notices that the screening of $\psi_b-\psi_0(z)$ is indeed much stronger than the decay experienced by $V_y(z)$. Then, for larger values of $b$ in Fig.~\ref{prchvsb}.(b) where the solvation energy $V_y(z)$ significantly decays, $k_y$  and $\Pi_y$ change their trend and begin to slowly increase towards $k_c$  and $\Pi_c$, respectively.

It is seen in Fig.~\ref{prchvsb}.(b) that for large enough $b$, the extra ionic exclusion effect associated with core interactions survives for all values of $\ell_y$ and $\e_m<\e_w$. More precisely, by comparing at fixed $b$ the curves for various $\ell_y$, one notices that an increase of the coupling parameter of the Yukawa potential that enhances the magnitude of the potential $V_y(z)$ amplifies the effect in question. Then, because a stronger dielectric discontinuity adds to the difference between the density of bulk and pore particles, the reduction of $\e_m$ increases the solvation energy $\ell_y(\kappa_{yb}-\kappa_y)$ in Eq.~(\ref{PMFslit2}), thus equally amplifying the additional ionic rejection effect.

We finally illustrate the interpolation between the two screening regimes by displaying in Fig.~\ref{prchvsb2}.(b) the competition between core collision effects and solvation forces as a function of $\rho_b$. One sees in this figure that in the regime of moderately screened core interactions (the curve for $b\ell_B=2.5$), the domination of solvation forces by core collisions leads to a partition function and an interplate pressure that rapidly increases with $\rho_b$. In this regime, the pressure remains less attractive than the pressure for the Coulomb liquid, until it becomes purely repulsive at $\rho_b\simeq 1$ M (see the inset). In the second regime of highly screened core interactions where repulsive solvation forces dominate core collision effects, one ends up with a partition coefficient that decays with increasing bulk ion concentration. Furthermore, $\Pi_y$ remains significantly more attractive than $\Pi_c$. It is also important to note that interestingly, the additional ionic exclusion phenomenon associated with excluded volume effects comes into play at significantly low bulk concentrations, for example at $\rho_b> 0.1$ M in the case $b\ell_B=5$.

It is known that the range of core interactions, or more precisely the exponential tail of the repulsive Yukawa potential is intimately related with the electronic structure of ions. Hence, the strong dependence of the ionic partition coefficients and the interplate pressure on the range of the repulsive Yukawa potential indicates these interactions as an important ion specific effect.

\begin{figure}
(a)\includegraphics[width=1.1\linewidth]{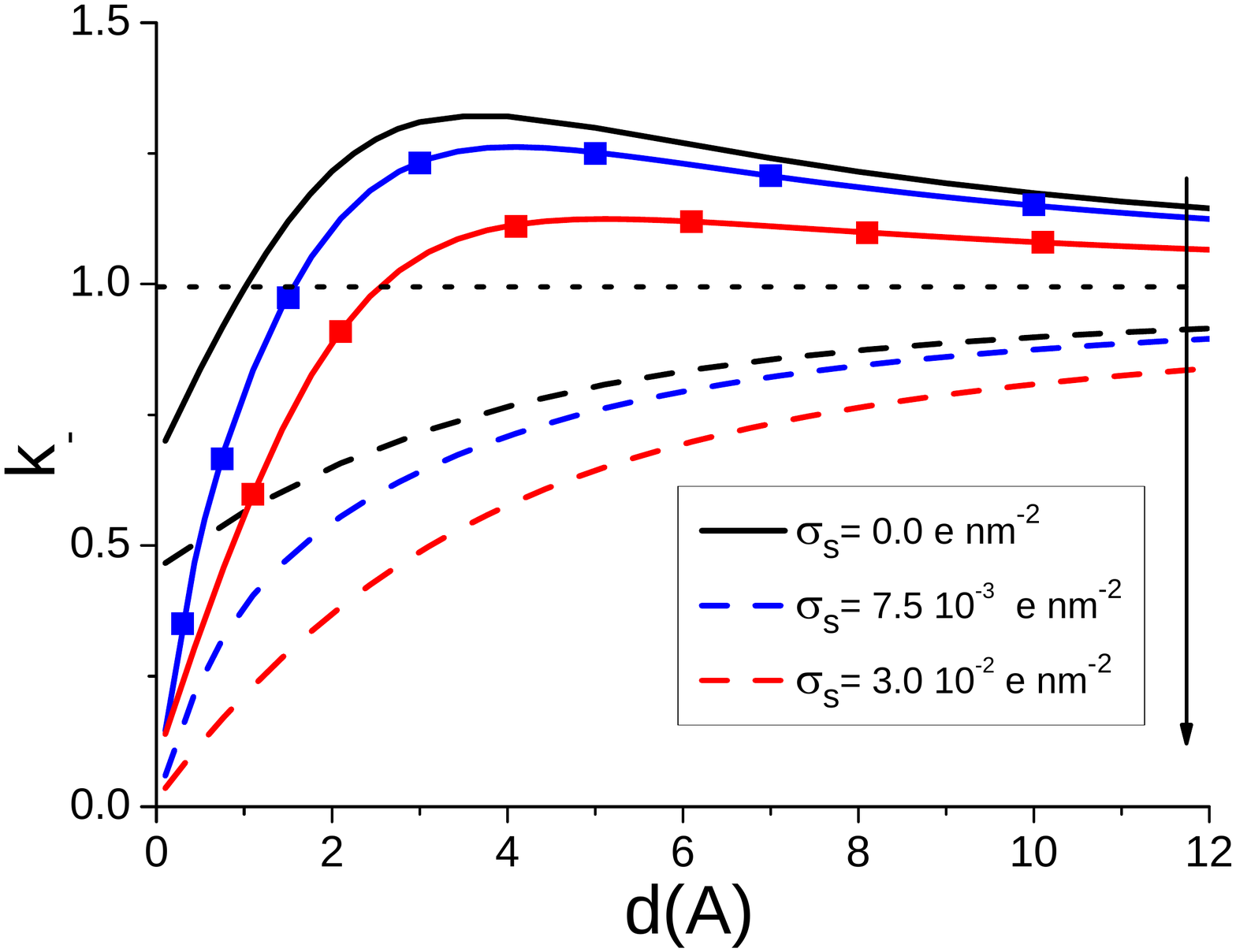}
(b)\includegraphics[width=1.1\linewidth]{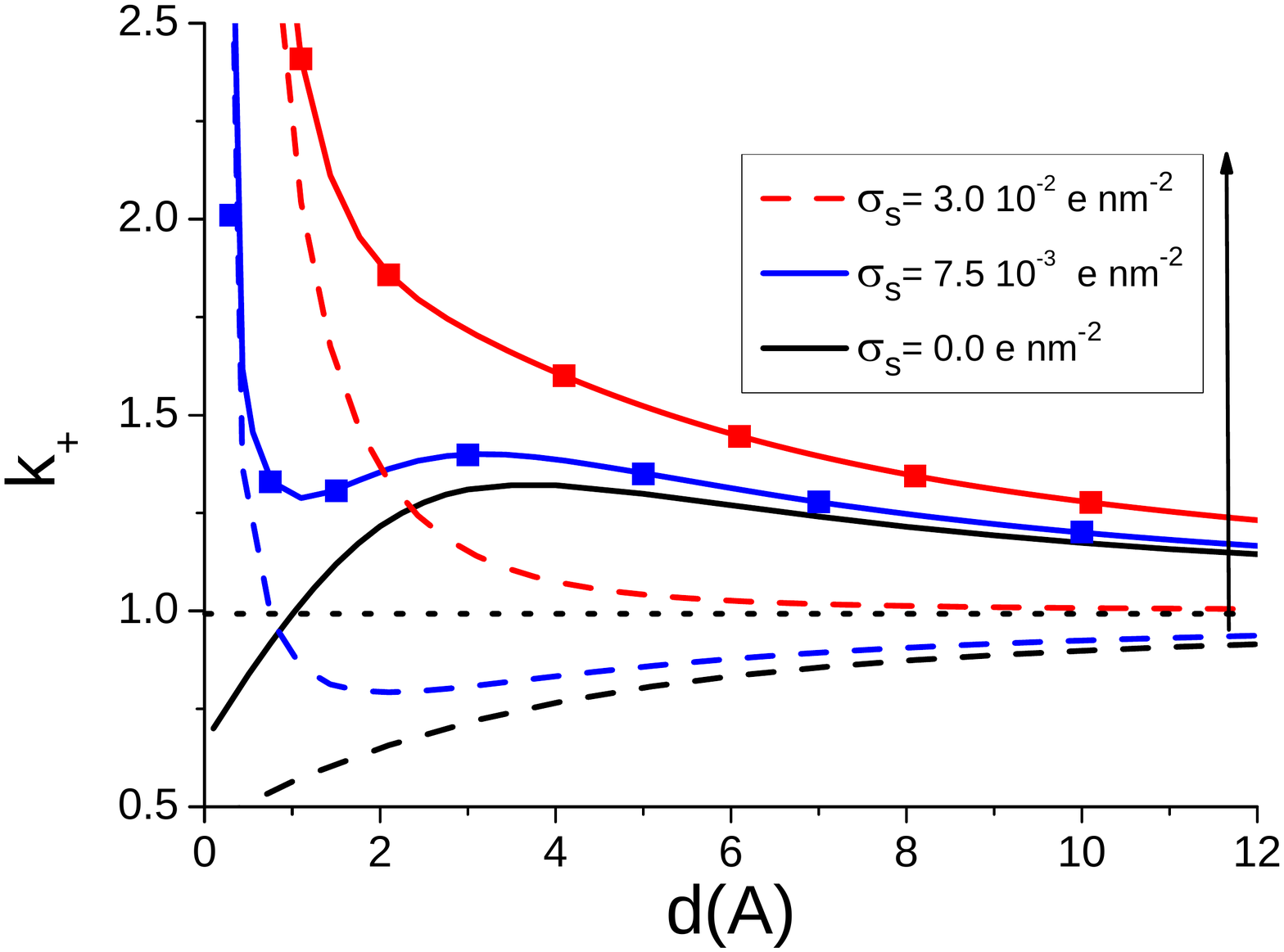}
(c)\includegraphics[width=1.1\linewidth]{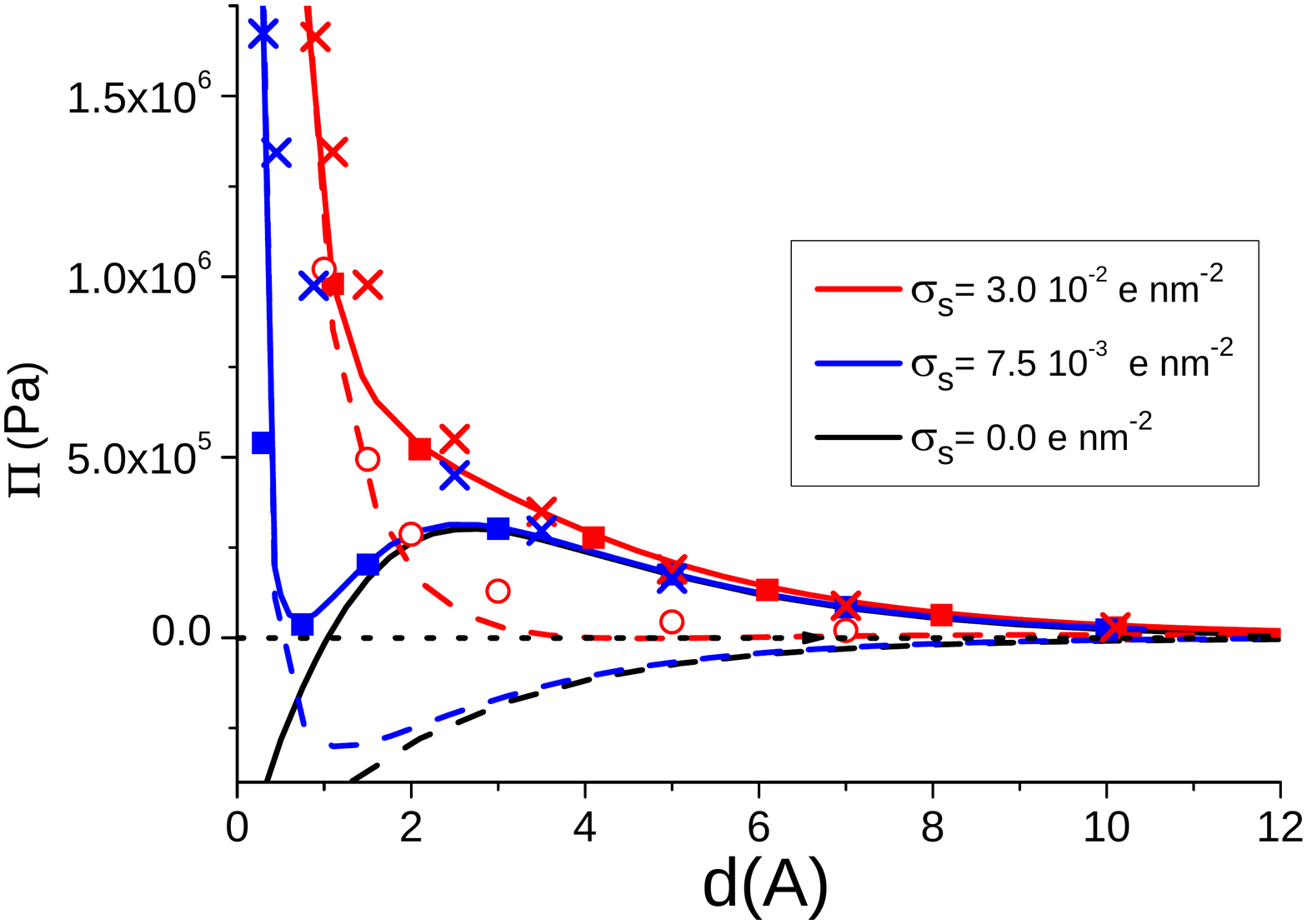}
\caption{(Color online) Partition coefficients of (a) coions and (b) counterions, and (c) the interplate pressure against the pore size for $\rho_b=0.5$ M and $\e_m=\e_w$. Dashed lines are for the Coulomb liquid ($\ell_y=0$) and solid lines correspond to the Yukawa liquid with $\ell_y=\ell_B$ and $b=2/\ell_B$. The curves are obtained from RVS and the squares mark the prediction of GVS. The circles and crosses illustrate the MF level pressure for the Coulomb and Yukawa fluids, respectively.}
\label{PrCh4}
\end{figure}
\begin{figure}
\includegraphics[width=1.1\linewidth]{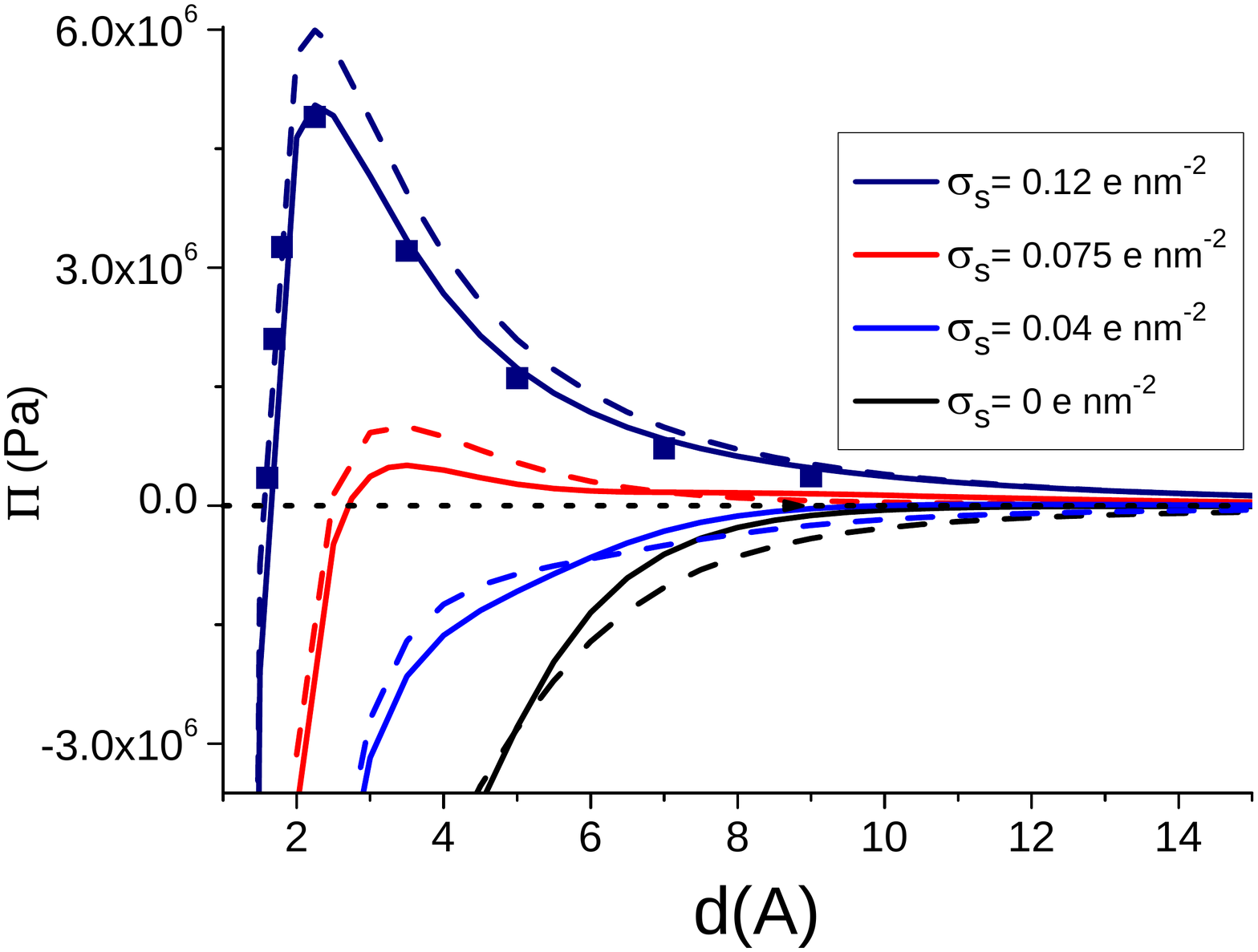}
\caption{(Color online) Interplate pressure against the pore size for $\e_m=2$, $\rho_b=0.5$ M and several values of $\sigma_s$. The curves are obtained from RVS and the squares  mark the prediction of GVS.}
\label{PrCh5}
\end{figure}

\subsection{Excluded volume effects in charged pores}

We investigate in this part the effect of a finite surface charge $\sigma_s$ on the distribution of Yukawa particles and the behaviour of the interplate pressure. In order to simplify the numerical task, the pressure and the partition coefficients will be computed within the RVS equations~(\ref{eqvarSC3})-(\ref{eqvarEX3}) and the results will be compared with the numerical solution of the GVS equations~(\ref{eqvarEX1})-(\ref{eqvarSC2}).

Fig.~\ref{PrCh4} illustrates the ionic partition coefficients and the net pressure for $\e_m=\e_w$, $\ell_y=0$ and $\ell_y=\ell_B$, and various values of $\sigma_s$. First of all, one notices for the Coulomb and the Yukawa liquids the close correlation between the partition coefficient of counterions and the pressure curves. Then, it is seen that in the absence of core interactions, the ionic selection of the charged pore is characterized by two regimes~\cite{PRE}. In small pores, one is in the GCE (good coion exclusion) regime where the ionic penetration is mainly fixed by the surface charge. Within this regime where $\gamma\gg \Gamma$, Eq.~(\ref{partcoDON}) yields $k_+\simeq\gamma\propto d^{-1}$ and $k_-\ll k_+$. In other words, there is an almost total coion exclusion and the counterion density rapidly drops with increasing pore size. For the weakest surface charge in this figure ($\sigma_s=7.5$ $10^{-3}$ $\mbox{nm}^{-2}$), this fast drop of the counterion density leads to a decrease of the initially positive pressure that becomes attractive at a characteristic pore size. In the opposite regime of large size pores where repulsive electrostatic solvation forces take over the electrostatic field created by the surface charge, the system behaves like a neutral pore. Namely, with increasing $d$, the density of both species  converge towards their bulk value and the attractive pressure decay in a monotonous way. Moreover, because the surface charge induced electric field is the only force that survives at the MF level, we show in Fig.~\ref{PrCh4}.(c) that the MF pressure agrees with the variational one exclusively in the GCE regime.

As seen in Fig.~\ref{PrCh4}, for weakly charged pores (the curve with $\sigma_s=7.5\times10^{-3}$ $\mbox{nm}^{-2}$), the inclusion of core interactions considerably complicates the picture described above. The neutral pore regime being now characterized by the particle packing effect, the pressure curves exhibit an oscillatory shape. Namely, in the transition regime between the GCE and neutral pore limits where core collisions driving the Yukawa charges into the pore as well as solvation forces start to dominate the electrostatic force induced by the surface charge, the counterion density and the pressure curve change their trend and begin to rise, until the characteristic pore size where $\Pi$ reaches the repulsive peak observed in the part on neutral pores (see Fig.~\ref{prch1}). Beyond this turning point, $k_\pm$ and $\Pi$ start to decay with the weakening of the particle packing effect. For stronger surface charges (e.g. the case $\sigma_s=3.0\times10^{-2}$ $\mbox{nm}^{-2}$), the GCE regime dominates up to larger pore sizes where solvation forces considerably weaken. Consequently, the oscillatory shape of the pressure curve is suppressed. In this case, excluded volume effects simply increase the pore density of both species and the magnitude of the interplate pressure. Furthermore, we show that the MF pressure agrees with the variational prediction in the GCE limit as well as the large $d$ (or the neutral pore) regime characterized by the ionic packing effect, whereas a discrepancy takes place between these regimes where correlation effects carried by solvation forces become significant.

Fig.~\ref{PrCh5} illustrates the behavior of the interplate pressure in the presence of a dielectric discontinuity with $\e_m=2$. It is seen that while increasing the surface charge, the ionic packing regime associated with $\Pi_y>\Pi_c$ is gradually reduced, until it dies out at large surface charges (i.e. $\sigma_s\gtrsim7.0\times10^{-2}$ $\mbox{nm}^{-2}$) where a pronounced repulsive peak takes place. In this regime of strong surface charges where the pore densities of the Coulomb and Yukawa liquid are very close, the difference between $\Pi_y$ and $\Pi_c$ is mainly due to the contribution from the core collisions to the outer pressure $\Pi_b$. We also show in Figs.~\ref{PrCh4} and~\ref{PrCh5}. that the restricted variational scheme agrees very well with the general one over a broad range of $\sigma_s$, and in the absence as well as in the presence of a dielectric discontinuity.

\section{Conclusions and discussion}

In this article, we analyzed the interaction force between two planar walls that confine a charged Yukawa liquid. We investigated the modifications of the predictions of the vdW and more elaborated theories~\cite{netzvdw,hatlo} by core interactions between the particles. To this aim, we used a recently developed self consistent calculation scheme~\cite{jstat} and also introduced a new restricted variational procedure that simplifies the numerical task of the former one. Both methods allow to take into account in a self consistent way the coupling between the surface charge induced electric field, image forces, and pore modified core interactions. We also developed the MF theory of the model in order to derive simple expressions for the MF level pressure that enabled us to identify the MF regime of the system.

In the first part, we considered the case of neutral Yukawa particles in a slit pore. We showed that due to core collisions within the bulk reservoir that drive the Yukawa particles into the pore, a net particle adsorption takes place. This ionic packing within the pore yields in turn a purely repulsive pressure.

In the second part, we studied the coupling of electrostatic and core interactions between Yukawa particles of finite charge in a neutral pore. It was shown that for very small pores without a dielectric discontinuity, the ionic rejection driven by repulsive solvation interactions leads to an attractive force between the pore walls for both Coulomb and Yukawa particles. With increasing pore size, core interactions that push the Yukawa particles into the pore leads to an ionic excess within the pore, and changes the sign of the pressure that becomes repulsive. It was also shown that the MF regime of the theory where the analytical expression for the interplate pressure agrees with the variational one corresponds to large pore sizes, e.g. $d>4$ {\AA} for $\rho_b=0.5$ M.

For small pore sizes and a strong dielectric discontinuity where the pore is in the dilute regime, core interactions contribute exclusively to the outer pressure mediated by the bulk particles, thereby increasing the magnitude of the attractive pressure. For larger pores where the magnitude of core collision effects become comparable with image forces, the accumulation of ions into the pore makes the pressure of the Yukawa liquid less attractive than the pressure associated with the Coulomb liquid. We also showed that for concentrated electrolytes ($\rho_b\gtrsim 1$ M), the interplate pressure may even become purely repulsive despite the strongly attractive vdW contribution.

In the third part of the article, we analyzed the role of the range of core interactions in the ionic rejection mechanism and the interaction force between the pore walls. We showed that the physics of the system can be roughly split into two screening regimes. The ionic packing effect discussed in the second part corresponds to the regime of moderately screened core interactions. In the second regime of strongly screened Yukawa potential, solvation interactions associated with core interactions result in an additional ionic exclusion from the pore, which in turn leads to an amplification of the magnitude of the attractive pressure. This unexpected effect that was shown to take place at quite low electrolyte concentrations calls for experimental verification. Furthermore, the high sensitivity of the interaction force between the plates to the range of core interactions suggests core-core collision effects as an ion specificity considerably more complex than a simple excluded volume effect induced by the packing fraction of charges.

The fourth part dealt with excluded volume effects in charged pores. We showed that for very weakly charged pores without a dielectric discontinuity, the intermediate regime between the GCE (small $d$) and the ionic packing (large $d$) regimes is characterized by oscillatory pressure curves. The oscillatory shape of the pressure is rapidly suppressed with an increase of the surface charge or the dielectric discontinuity. Furthermore, it was shown that the MF pressure agrees with the variational one exclusively in the GCE and ionic packing regimes. In the presence of a dielectric discontinuity, while increasing the surface charge, the ionic packing regime gradually disappears and a large repulsive peak sets in at small interplate separations. In this range of pore sizes, core interactions solely increase the external pressure $\Pi_b$, thereby decreasing the magnitude of the net pressure. We can conclude by noting that the significant repulsive contribution from core collisions for neutral and weakly charged pores suggest excluded volume effects as an important ingredient for the stabilization of colloidal molecules in electrolyte solutions.

It was also shown that over a broad range of surface charge, dielectric discontinuity, and bulk concentration, the restricted variational equations show a very good agreement with the prediction of the more general equations for the partition coefficients and the interplate pressure. As stressed in the text, the restricted variational scheme can be helpful if one wishes to consider the excluded volume effects in more complicated geometries, such as cylindrical ion channels, whose importance in biophysics is well established~\cite{Hille,Bob}. It is also important to note that non-linear equations similar in form to Eq.~(\ref{eqvarSC3}) are frequently used by the nanofiltration community in water purification and desalination processes~\cite{yarosch,Szymczyk}. Hence, the restricted self-consistent scheme presents itself as a practical method to check the importance of excluded volume effects in the artificial nanofiltration process, where very concentrated salt solutions are used in experiments. Finally, the idea used in the derivation of the restricted equations could be followed as well in order to take into account additional non-electrostatic particle interactions in various pore geometries.

It is important to emphasize that the present theory is based on the dielectric continuum formulation of electrostatics, where the solvent molecules are considered as a simple dielectric background that locally renormalizes  the dielectric permittivity of the air, rather than polar molecules subject to electrostatic interactions present in the system. An explicit electrostatic modeling of solvent molecules presents itself as a non-trivial problem that we are currently investigating. Furthermore, we note that for the concentrated solvent system investigated in this work (solvent density $\rho_w=55$ M and bulk dielectric permittivity $\e_w=78$), the core-core collisions between solvated ions are associated with the excluded volume of hydrated ion size (i.e. an ion surrounded by solvent molecules) rather than the bare ion size. Investigating core collisions in low density solvents where ionic solvation is not perfect may require a consideration of solvent molecules and ions as separate entities. However, the complications resulting from both issues discussed in this paragraph are beyond the scope of this article.

As discussed in Ref.~\cite{jstat}, the present theory is a first order approach that has its limitations. First of all, it is based on a generalized Onsager-Samaras theory characterized by uniform screening parameters. Moreover, the closure relations~(\ref{eqvarEX1})-(\ref{eqvarSC2}) are derived from a first order cumulant expansion. In view of these limitations, we considered exclusively the submolar concentration regime of monovalent ions and weak surface charges, namely $\sigma_s\leq0.12$ $\mbox{nm}^{-2}$. This upper boundary for the surface charge corresponds to an electrostatic coupling parameter $\Xi=0.37$, i.e. the weak coupling regime. We are currently working on the relaxation of the generalized Onsager-Samaras approximation, which will enable us to estimate the error induced by this approximation. To our knowledge, MC simulations for Yukawa charges in slit pores are still missing. When these simulation data become available, it will be necessary to test the accuracy of the theoretical tools presented in this article.
\\
\acknowledgements This work has been supported in part by The Academy of Finland through its COMP CoE and NanoFluid grants.
\smallskip

\appendix
\section{Numerical solution of the self-consistent equations}
\label{numsol}

We explain in this appendix the numerical solution of the self consistent equations~(\ref{eqvarEX1})-(\ref{eqvarSC2}) and~(\ref{eqvarEX3})-(\ref{eqvarSC3}) by iteration. First of all, by using the bulk values $\kappa_c^{in}=\kappa_{DH}$ and $\kappa_y^{in}=\kappa_{yb}$ in the potentials $V_y(z)$ and $V_c(z)$, the Eqs.~(\ref{eqvarEX1}) and~(\ref{eqvarEX2}) should be solved either with Wolfram Mathematica 7 software or with a Fortran code. Then, the obtained potential profiles $\phi_0(z)$ and $\psi_0(z)$ are injected together with $\kappa^{in}_c$ and $\kappa_y^{in}$ into Eq.~(\ref{eqvarSC1}) in order to obtain the updated value $\kappa_c^{out}$.  The potential profiles and $\kappa_c^{out}$ are finally put into Eq.~(\ref{eqvarSC1}) to obtain the new value $\kappa_y^{out}$. The updated screening parameters are then substituted into the equations~(\ref{eqvarEX1}) and~(\ref{eqvarEX2}) in the second iteration to obtain the new potential profiles, and the cycle is repeated until self-consistency is achieved. The solution to the RVS equations~(\ref{eqvarSC3})-(\ref{eqvarEX3}) can be found with the same iterative scheme.

The equations~(\ref{eqvarEX1}) and~(\ref{eqvarEX2}) can be solved with Wolfram Mathematica 7 software for  weak surface charges, submolar concentrations, and pore sizes below 1 nm. For model parameters where the numerical solver of the software fails, we used a Fortran code with a 4th order Runge-Kutta integrator. In the Fortran code, a shooting algorithm that allows to find the surface potentials $\psi_0(0)$ and $\phi_0(0)$ satisfying the mixed boundary conditions~(\ref{boundphi2})-(\ref{boundpsi3}) was used. The shooting method consists in integrating first these differential equations from the surface at $z=0$ until the mid-pore with an arbitrary initial surface potential. Depending on the sign of the derivative of the potential in the mid-pore, the initial surface potential is updated with a lower or a higher value and the numerical integration is reiterated with the new boundary values. This procedure is repeated until the derivative of the potential vanishes in the middle of the pore. We finally note that the choice of the updated values for the surface potentials was done with a dichotomy algorithm in order to minimize the number of iteration of the numerical integration.

\section{Electrostatic and Yukawa kernels}
\label{appendixCoef}

In this appendix, we report the electrostatic and Yukawa potentials for the slit pore geometry depicted in Fig.~\ref{sketch}. These potentials are defined as the inverse of the operators in Eqs.~(\ref{DH1}) and~(\ref{DH2}). They can be obtained with a single calculation as explained in Ref.~\cite{jstat}, i.e. by inverting the generalized DH equation
\be\label{DH3}
\left[-\nabla(\epsilon(\br)\nabla)+\epsilon(\br)\kappa^2(\br)\right]U(\br,\br')=\lambda\delta(\br-\br')
\ee
with the piecewise dielectric permittivity and the inverse screening length given by
\bea\label{pwe}
\e(z)&=&\e_>\theta(z)\theta(d-z)+\e_<[\theta(-z)+\theta(z-d)]\hspace{0.5cm}\\
\label{pwk}
\kappa(z)&=&\kappa_<[\theta(-z)+\theta(z-d)]\nonumber\\
&&+\kappa_>\theta(z)\theta(d-z).
\eea
The free parameter $\lambda$ will be fixed in the end in order to recover $v_0(\br,\br')$ and $w_0(\br,\br')$ from $U(\br,\br')$. The solution of Eq.~(\ref{DH3}) with Eqs.~(\ref{pwe}) and~(\ref{pwk}) is rather trivial and can be found for example in Ref.~\cite{netzvdw}. The kernel is composed of a bulk and an anisotropic part, that is
\be\label{kernel0}
U(\br,\br')=\frac{\lambda}{4\pi\e(z)}\frac{e^{-\kappa(z)|\br-\br'|}}{|\br-\br'|}+\delta U(\br,\br').
\ee
The derivation of the results presented in this article requires exclusively the knowledge of the Green's function evaluated at the same point, which reads
\be\label{Ker1}
\delta U(\br'=\br)=-\frac{\lambda}{4\pi\e_<}\int_0^\infty\frac{\mathrm{d}kk\Delta}{\rho_<}\frac{1-e^{-2\rho_>d}}
{1-\Delta^2e^{-2\rho_>d}}e^{2\rho_<z}
\ee
if $z\leq0$,
\bea\label{Ker2}
\delta U(\br'=\br)&=&-\frac{\lambda}{4\pi\e_>}\int_0^\infty\frac{\mathrm{d}kk\Delta}{\rho_>}\\
&&\times\frac{e^{-2\rho_>z}+e^{-2\rho_>(d-z)}+2\Delta e^{-2\rho_>d}}{1-\Delta^2e^{-2\rho_>d}}\nonumber
\eea
if $0\leq z\leq d$ and
\be\label{Ker3}
\delta U(\br'=\br)=-\frac{\lambda}{4\pi\e_<}\int_0^\infty\frac{\mathrm{d}kk\Delta}{\rho_<}\frac{1-e^{-2\rho_>d}}
{1-\Delta^2e^{-2\rho_>d}}e^{2\rho_<(d-z)}
\ee
if $z\geq d$, where we have defined
\be
\Delta=\frac{\rho_>-\eta\rho_<}{\rho_>+\eta\rho_<},
\ee
$\rho_\lessgtr=\sqrt{k^2+\kappa_\lessgtr^2}$ and $\eta=\e_</\e_>$.

First, by setting $\kappa_<=0$, $\kappa_>=\kappa_c$, $\e_<=\e_m$, $\e_>=\e_w$, $\lambda=4\pi\e_>\ell_B$ and defining $\rho_c=\sqrt{\kappa_c^2+k^2}$,
\be
\Delta_c=\frac{\rho_c-\eta_c k}{\rho_c+\eta_ck} ,\hspace{5mm}\eta_c=\e_m/\e_w,
\ee
one recovers from Eqs.~(\ref{Ker1})-(\ref{Ker3}) the variational electrostatic potential $v_0(z)$.

Second,  by setting $\kappa_<=b$, $\kappa_>=\kappa_y$, $\e_<=\e_>$, $\lambda=4\pi\e_>\ell_y$, and defining $\rho_m=\sqrt{b^2+k^2}$, $\rho_y=\sqrt{\kappa_y^2+k^2}$ and
\be
\Delta_y=\frac{\rho_y-\rho_m} {\rho_y+\rho_m},
\ee
Eqs.~(\ref{Ker1})-(\ref{Ker3}) yield the variational Yukawa potential $w_0(z)$.

\section{Computation of the excess Grand potential with the charging method}
\label{appendixExc}

We explain in this appendix the derivation of the excess Grand potential with the charging procedure. The computation will be carried out in terms of the general kernel $U(\br,\br')$ (see Appendix~\ref{appendixCoef}) and the fluctuating potential $\varphi$ in order to recover from the final result the excess potential of electrostatic and Yukawa interactions.

The charging method consists in reexpressing the gaussian part of the Grand potential in terms of integrals over auxiliary charging parameters $\eta$ and $\xi$, that is
\label{appendixChr}
\bea
\Omega_0&=&-\ln\int\mathcal{D}\varphi\;e^{-\int\frac{\mathrm{d}\br}{2\lambda}\e(\br)\left[(\nabla\varphi)^2
+\kappa(\br)^2\varphi^2\right]}\\
&=&-\int_0^1\mathrm{d}\xi\frac{\mathrm{d}}{\mathrm{d}\xi}\ln\int\mathcal{D}\varphi\;e^{-\int\frac{\mathrm{d}\br}{2\lambda}\e(\br)
\left[(\nabla\varphi)^2+\kappa_\xi^2(\br)\varphi^2\right]}\nonumber\\
&&-\int_0^1\mathrm{d}\eta\frac{\mathrm{d}}{\mathrm{d}\eta}\ln\int\mathcal{D}\varphi\;e^{-\int\frac{\mathrm{d}\br}{2\lambda}\e(\br)
\left[(\nabla\varphi)^2+\eta\kappa_<^2(\br)\varphi^2\right]}\nonumber\nonumber\\
&&-\ln\int\mathcal{D}\varphi\;e^{-\int\frac{\mathrm{d}\br}{2\lambda}\e(\br)(\nabla\varphi)^2}\nonumber\\
&=&\Omega_{01}+\Omega_{02}+\Omega_{vdW}
\eea
where we defined the screening parameter $\kappa_\xi^2(\br)=\kappa_<^2+\xi\left[\kappa(\br)^2-\kappa_<^2\right]$. The unscreened vdW energy reads~\cite{netzvdw}
\be
\Omega_{vdW}=\frac{S}{4\pi}\int_0^\infty\mathrm{d}kk\ln\left(1-\Delta_0^2e^{-2kd}\right).
\ee
The remaining two terms $\Omega_{01}$ and $\Omega_{02}$ can be calculated by evaluating the derivatives with respect to $\eta$ and $\xi$, which yields
\bea\label{om1}
\Omega_{01}&=&S\frac{\kappa_>^2-\kappa_<^2}{2\lambda}\e_2\int_0^d\mathrm{d}z\int_0^1\mathrm{d}\xi U\left(\br'=\br;\kappa(\br)\to\kappa_\xi(\br)\right)\\
&=&\frac{Sd}{4\pi}(\kappa_>^2-\kappa_<^2)\int_0^\infty\frac{\mathrm{d}kk}{\rho_>+\rho_<}\nonumber\\
&&+\frac{S}{4\pi}\int_0^\infty\mathrm{d}kk\ln\frac{(\rho_>+\eta\rho_<)^2}{(\eta+1)^2\rho_>\rho_<}\nonumber\\
&&+\frac{S}{4\pi}\int_0^\infty\mathrm{d}kk\left[\ln\left(1-\Delta^2e^{-2\rho_>d}\right)-
\ln\left(1-\Delta_0^2e^{-2\rho_<d}\right)\right]\nonumber\\
\label{om2}
\Omega_{02}&=&S\frac{\kappa_<^2}{2\lambda}\int_{-\infty}^\infty\e(z)\mathrm{d}z\int_0^1\mathrm{d}\eta U\left(\br'=\br;\kappa(\br)\to\eta\kappa_<\right)\\
&=&\frac{S}{4\pi}\int_0^\infty\mathrm{d}kk\left[\ln\left(1-\Delta_0^2e^{-2\rho_<d}\right)-
\ln\left(1-\Delta_0^2e^{-2kd}\right)\right],\nonumber
\eea
where we used the kernel Eqs.~(\ref{Ker1})-(\ref{Ker3}) and analytically performed the integrals over $z$, $\eta$ and $\xi$. We also introduced
\be
\Delta_0=\frac{\e_w-\e_m}{\e_w+\e_m}.
\ee
Although the first and second integrals in Eq.~(\ref{om1}) are UV-divergent, this divergence is artificial and disappears when we substract the correction terms arising in the cumulant expansion in Eq.~(\ref{grandPot}), i.e.
\be
\frac{\Delta\Omega_0}{S}\equiv\frac{\Omega_0}{S}-\frac{\kappa_>^2-\kappa_<^2}{2\lambda}\e_>\int_0^dU\left(\br'=\br;\kappa(\br)\right)
\ee
which reads
\bea\label{DelOm}
&&\frac{1}{S}\Delta\Omega_0(\kappa_<,\kappa_>,\eta)=\frac{d}{24\pi}(\kappa_>-\kappa_<)(\kappa_>^2+\kappa_>\kappa_<-2\kappa_<^2)\nonumber\\
&&+\frac{\Delta_0}{16\pi}(\kappa_>^2-\kappa_<^2)+\frac{\kappa_<^2}{8\pi}\ln\frac{(\eta+1)^2\kappa_<\kappa_>}
{(\eta\kappa_<+\kappa_<)^2}\nonumber\\
&&+\int_0^\infty\frac{\mathrm{d}kk}{4\pi}\ln\left(1-\Delta^2e^{-2\rho_>d}\right)\nonumber\\
&&-\frac{\kappa_>^2-\kappa_<^2}{8\pi}\int_0^\infty\mathrm{d}kk\frac{\Delta}{\rho_>^2}\frac{\Delta^2+2d\rho_>\Delta-1}
{1-\Delta^2e^{-2\rho_>d}}e^{-2\rho_>d}\nonumber.
\eea
The Coulombic and Yukawa parts $\Delta\Omega_{0\phi}$ and $\Delta\Omega_{0\psi}$ directly follow from Eq.~(\ref{DelOm}) by making the identification explained in Appendix~\ref{appendixCoef}, that is
\bea
&&\Delta\Omega_{0\phi}=\Delta\Omega_0(\kappa_<=0,\kappa_>=\kappa_c,\eta_c)\\
&&\Delta\Omega_{0\psi}=\Delta\Omega_0(\kappa_<=b,\kappa_>=\kappa_y,\eta=1).
\eea

\end{document}